\documentclass[12pt]{article}
\pdfoutput=1    

\usepackage{achemso}
\setkeys{acs}{articletitle=true,maxauthors=20,chaptertitle=true}  

\usepackage{times}

\title{Organic molecules as origin of visible-range single photon emission from hexagonal boron nitride and mica}

\author
{Michael Neumann$^{1,2\dag\ast}$, Xu Wei$^{1,3\dag}$, Luis Morales-Inostroza$^{4}$, Seunghyun Song$^{1,5}$,\\
Sung-Gyu Lee$^{1,3}$, Kenji Watanabe$^{6}$, Takashi Taniguchi$^{7}$, Stephan G{\"o}tzinger$^{4,8,9}$,\\
Young Hee Lee$^{1,3\ast}$\\
\\
\normalsize{$^{1}$Center for Integrated Nanostructure Physics, Institute for Basic Science (IBS),}\\
\normalsize{Suwon 16419, Republic of Korea}\\
\normalsize{$^{2}$Sungkyunkwan University, Suwon 16419, Republic of Korea}\\
\normalsize{$^{3}$Department of Energy Science, Sungkyunkwan University, Suwon 16419, Republic of Korea}\\
\normalsize{$^{4}$Max Planck Institute for the Science of Light, D-91058 Erlangen, Germany}\\
\normalsize{$^{5}$Department of Electronics Engineering, Sookmyung Women's University,}\\
\normalsize{Seoul 04310, Republic of Korea}\\
\normalsize{$^{6}$Research Center for Functional Materials, National Institute for Materials Science,}\\
\normalsize{1-1 Namiki, Tsukuba 305-0044, Japan}\\
\normalsize{$^{7}$International Center for Materials Nanoarchitectonics,}\\
\normalsize{National Institute for Materials Science, 1-1 Namiki, Tsukuba 305-0044, Japan}\\
\normalsize{$^{8}$Department of Physics, Friedrich-Alexander University Erlangen-N{\"u}rnberg (FAU),}\\
\normalsize{D-91058 Erlangen, Germany}\\
\normalsize{$^{9}$Graduate School in Advanced Optical Technologies (SAOT),}\\
\normalsize{Friedrich-Alexander University Erlangen-N{\"u}rnberg (FAU), D-91052 Erlangen, Germany}\\
\\
\normalsize{$^\dag$These authors contributed equally to this work.}\\
\normalsize{$^{\ast}$Corresponding authors. E-mail: mneumann@skku.edu (M.N.); leeyoung@skku.edu (Y.H.L.)}
}

\date{}

\usepackage{graphicx}

\usepackage{amssymb}

\usepackage[version=3]{mhchem} 

\newcommand*{\degC}{$\!^\circ$C}

\newcommand*{\invcm}{cm$^{-1}$}

\begin{document}

\baselineskip18pt

\maketitle

\section*{Abstract}

The discovery of room-temperature single-photon emitters (SPEs) hosted by two-dimen\-sional hexagonal boron nitride (2D hBN)
has sparked intense research interest. Although emitters in the vicinity of 2 eV have been studied extensively, their
microscopic identity has remained elusive. The discussion of this class of SPEs has centered on point defects in the hBN
crystal lattice, but none of the candidate defect structures have been able to capture the great heterogeneity in emitter
properties that is observed experimentally. Employing a widely used sample preparation protocol but disentangling several
confounding factors, we demonstrate conclusively that heterogeneous single-photon emission $\sim 2$ eV associated with hBN
originates from organic molecules, presumably aromatic fluorophores. The appearance of those SPEs depends critically on the
presence of organic processing residues during sample preparation, and emitters formed during heat treatment are not located
within the hBN crystal as previously thought, but at the hBN/substrate interface. We further demonstrate that the same class of
SPEs can be observed in a different 2D insulator, fluorophlogopite mica.

\noindent \textbf{Keywords:} single photon emission, hexagonal boron nitride, fluorophlogopite mica, organic
processing residue, polycyclic aromatic hydrocarbons

\section*{Introduction}

Single photon emitters (SPEs) hosted in crystalline solids have received intense attention as promising elements for quantum
information processing and metrology applications \cite{Zhang2020a}. Layered van der Waals (vdW) materials that host SPEs are
particularly desirable due to their compatibility with standard lithographic processing techniques. The discovery that
hexagonal boron nitride (hBN) features bright, photostable, and robust room-temperature SPEs emitting in the visible wavelength
range \cite{Tran2016a} has motivated intense research efforts devoted to the exploration of these emitters and their
integration into nanophotonic device structures \cite{Toth2019,Caldwell2019,Sajid2020a}.

A central open question in the field is the microscopic identity of the emission centers, knowledge of which would enable the
development of truly controlled SPE preparation protocols. Here, we focus on the most widely studied type of SPE, which emits
in the vicinity of 2 eV and exhibits striking inter-emitter variability. While numerous defect structures in the
hBN crystal lattice have been proposed as candidates, the SPE's identity has remained unresolved, primarily due
to the great heterogeneity observed in emitters' behavior in this energy range, which renders an explanation in terms of one or
a few point defect types difficult.

SPEs can be produced using different starting materials, including both liquid-exfoliated
\cite{Tran2016b,Jungwirth2016a,Shotan2016,Kozawa2023} and mechanically cleaved hBN
\cite{Tran2016c,Chejanovsky2016,Choi2016,Xu2018,Fischer2021}, as well as hBN grown by chemical vapor deposition (CVD)
\cite{Tran2016a,Chejanovsky2016,Li2017,Stern2019,Mendelson2019,Mendelson2020,Li2021,Kozawa2023}. Sample preparation protocols
commonly contain an ``activation'' heat treatment step in inert atmosphere, typically at 850
\degC.\cite{Tran2016a,Tran2016b,Jungwirth2016a,Shotan2016,Tran2016c,Chejanovsky2016,Choi2016,Xu2018,Fischer2021,Kozawa2023} A
variety of approaches has been shown to activate or stabilize emission centers, including irradiation with electrons
\cite{Tran2016b,Choi2016}, ions \cite{Chejanovsky2016,Choi2016,Grosso2017,Mendelson2021}, intense laser pulses \cite{Choi2016},
as well as depositing hBN sheets on arrays of pillars \cite{Proscia2018} and etching with acids \cite{Chejanovsky2016} or
plasma \cite{Xu2018,Fischer2021}. These methods are frequently combined with activation heat treatment in an oxygen-free
atmosphere.

The reported physical properties of emission centers in hBN exhibit significant heterogeneity. The most frequently reported
type of PL spectrum is dominated by a sharp emission maximum, referred to in the literature as zero phonon line (ZPL). The
energy of this main emission peak often varies widely between emitters within the same sample, ranging between 1.6 and 2.2
eV.\cite{Tran2016b,Chejanovsky2016,Martinez2016,Shotan2016,Xu2018,Kozawa2023} It is often accompanied by one or several
emission maxima that are red-shifted by $150 \sim 170$
meV,\cite{Tran2016b,Martinez2016,Shotan2016,Li2017,Exarhos2017,Xu2018,Kozawa2023} commonly designated as phonon side bands
(PSB). However, the shapes of emission spectra can vary greatly \cite{Li2017}, even between emitters for which single-photon
emission has been demonstrated via antibunching in the second-order correlation function, $g^{(2)}(0) <
0.5$.\cite{Tran2016b,Exarhos2017} Fluorescence intensity is emitter-specific but can be very bright
\cite{Tran2016a,Tran2016b,Martinez2016,Grosso2017,Mendelson2019,Mendelson2020}, with excited-state lifetimes in the range of $1
- 10$ ns,\cite{Tran2016a,Tran2016b,Tran2016c,Jungwirth2016a,Chejanovsky2016,Schell2016,Grosso2017,Exarhos2017} although
additional states with significantly longer time constants have been reported as well
\cite{Tran2016b,Tran2016c,Chejanovsky2016,Martinez2016,Exarhos2017}. Emitters frequently exhibit polarized light absorption and
emission characteristics \cite{Tran2016a,Tran2016c} with great variability between emitters
\cite{Jungwirth2016a,Choi2016,Exarhos2017,Kozawa2023}, as well as variable degrees of fluorescence intermittency (``blinking'')
and spectral diffusion \cite{Martinez2016,Shotan2016,Li2017,Stern2019}. Properties of individual emitters are sensitive to
external parameters such as strain, hydrostatic pressure, and electric field
\cite{Grosso2017,Xue2018,Noh2018,Nikolay2019a,Mendelson2019,Xia2019}, again with significant emitter-to-emitter variability,
and a small subset of emission centers exhibits a response to magnetic fields
\cite{Exarhos2019,Mendelson2021,Chejanovsky2021,Stern2022}.

There have been numerous attempts to reproduce the measured emitter properties in theoretical calculations. The pioneering work
of Tran \textit{et al.} \cite{Tran2016a} proposed that, analogous to $\mathrm{N}_\mathrm{C} \mathrm{V}_\mathrm{C}$ color
centers that are the source of single-photon emission in diamond, visible emission from hBN samples originates from
$\mathrm{N}_\mathrm{B} \mathrm{V}_\mathrm{N}$ antisite complex defects in the crystal lattice
\cite{Tran2016a,Abdi2018,Xue2018,Noh2018}. Subsequent theoretical work has proposed additional candidates for single-photon
emission in the range of $\sim 2$ eV, such as the vacancy defects $\mathrm{V}_\mathrm{N}$ \cite{Li2017,Wang2018,Korona2019},
$\mathrm{V}_\mathrm{B}$ \cite{Li2017,Abdi2018,Wang2018}, and their clusters \cite{Li2017}, carbon substitutions including
$\mathrm{C}_\mathrm{B}$ and $\mathrm{C}_\mathrm{N}$ \cite{Jara2021,Auburger2021}, the complexes $\mathrm{C}_\mathrm{B}
\mathrm{V}_\mathrm{N}$ \cite{Tawfik2017,Abdi2018,Sajid2018,Noh2018,Sajid2020b}, $\mathrm{C}_\mathrm{N} \mathrm{V}_\mathrm{B}$
\cite{Mendelson2021} and $\mathrm{O}_\mathrm{B} \mathrm{V}_\mathrm{N}$ \cite{Noh2018}, and others
\cite{Weston2018,Xu2018,Turiansky2019}. There is significant disagreement between different works regarding the viability of
these defect candidates in accounting for the observed emission properties. The recent experimental demonstration that carbon
is involved in the formation of emission centers \cite{Mendelson2021,Lyu2020} eliminates many of the candidate defect
structures. However, the great variability of reported properties of SPEs complicates the explanation of the observed phenomena
in terms of one or few types of lattice point defects. The presence of a variable local dielectric environment, charge traps,
strain, or multiple charged defect states have been hypothesized to account for the observed spread of emitter properties
\cite{Tran2016b,Chejanovsky2016,Martinez2016,Jungwirth2016a,Shotan2016,Li2017,Grosso2017}. Recently it has been demonstrated
that at least four distinct types of emitters exist \cite{Hayee2020}. Nevertheless, no agreement on the identity of SPEs $\sim
2$ eV has emerged so far.

In this work, we demonstrate a qualitatively different mechanism of the single-photon emission $\sim 2$ eV from
hBN samples that is characterized by great inter-emitter variability. Namely, the emitters are aromatic
fluorophores that are formed under hBN sheets from organic processing residues during heat treatment in an
oxygen-free atmosphere, and they remain trapped at the hBN/substrate interface. We argue that polycyclic aromatic
hydrocarbon (PAH) molecules, which are also referred to as nanographene molecules, are the most likely candidates.

PAHs are a large family of organic compounds, a variety of which are produced when organic matter is exposed to high
temperature under oxygen-deficient conditions \cite{Norinaga2009,Viteri2019,Zhou2015}. Many PAH species exhibit fluorescence
\cite{Rieger2010}, and several that possess particularly favorable properties have been studied extensively as SPEs
\cite{Toninelli2021}. Given that the composition and distribution of organic residues are typically uncontrolled in hBN sample
preparation, and that heat treatment can lead to a large number of PAH reaction products, the scenario outlined here quite
naturally accommodates the great heterogeneity seen in the properties of emission centers in hBN samples (see also
Supplementary Note 1).

We note that besides the most extensively reported, highly variable SPEs discussed in this work, hBN also hosts several other
classes of emitters that exhibit uniform behavior, in particular very reproducible spectral shapes and emission energies,
mostly far from 2 eV, in the ultraviolet, blue, and infrared ranges. The scenario outlined in this work does not apply to these
qualitatively different emitters (see Supplementary Note 2 for discussion).

\section*{Results}

\subsection*{Appearance of SPEs depends on presence of organic residue}

We begin by demonstrating that the appearance of emission centers critically depends on the presence of organic residues during
sample preparation. Our protocol is designed to mimic widely used preparation methods, but with a simpler sample structure, and
with added steps for the controlled removal and reintroduction of organic impurities. Using high-quality bulk hBN as starting
material \cite{Watanabe2004}, we prepare specimens by mechanical exfoliation, which offers the benefit of a simple sample
geometry without restacked sheets, and allows us to use a cleaning protocol that includes a heating step in oxygen atmosphere
to fully remove all organic adhesive residues. As illustrated in Fig. 1a, we use two contrasting preparation methods that
enable us to disentangle the effects of heat treatment and exposure to organic impurities:

\begin{itemize}
\item In a control cycle, hBN samples undergo oxidative heat treatment for removal of organic contaminants. Directly after
    cool-down, they are subjected to the widely used activation heat treatment at 850 \degC\ in inert atmosphere.
\item In a `live' cycle, an additional step of sample immersion in organic solvent is inserted after oxidative cleaning and
    prior to activation heat treatment. This step facilitates the controlled introduction of organic impurities by solvent
    diffusion between hBN sheets and the substrate \cite{Gasparutti2020}.
\end{itemize}

The outcomes of these two protocols differ dramatically: After the thin hBN sheet shown in Fig. 1b undergoes the control cycle,
its photoluminescence (PL) intensity map exhibits only very weak background fluorescence, and no evidence of SPEs (Fig. 1c). In
contrast, when the same hBN sample subsequently undergoes a live cycle involving exposure to ethanol, the PL intensity map
features bright emission centers, accompanied by a weaker uniform PL background (Fig. 1d). These observations are not specific
to ethanol: Figure 1e-g shows another hBN sheet, for which the control treatment results in a dark PL intensity map without
emission centers, whereas the following live cycle involving immersion in the aromatic solvent anisole leads to numerous bright
emitters on a homogeneous PL background.

To exclude the scenario that the appearance of emission centers is merely the result of accumulated hBN crystal damage from
multiple subsequent heat treatments, we demonstrate that two consecutive control cycles, without sample exposure to organic
solvents, do not lead to the formation of emission spots (Fig 1h-j). Conversely, an hBN sample subjected only to a live cycle
with immersion in anisole, without a preceding control cycle, results in a PL map with many bright emission spots (Fig. 1k,l).

Our observations demonstrate conclusively that the generation of PL emission centers requires the presence of organic residues
during the activation heat treatment step in inert atmosphere. This finding is consistent with recent reports that have
provided evidence for the involvement of carbon in emitter formation \cite{Mendelson2021,Lyu2020}. However, while those
previous works interpreted the link between emitter appearance and the presence of carbon within the framework of point defects
in the hBN crystal lattice, for example $\mathrm{C}_\mathrm{B} \mathrm{V}_\mathrm{N}$, here we propose the qualitatively
different scenario of extrinsic, PAH-like fluorescent molecules as the emitter origin. We highlight the similarity between the
spectra of emitters in hBN and those of PAH molecules in Fig. 1m. Typical PL spectra of emission centers that appear in hBN
sheets after live cycles with exposure to ethanol or anisole (green, yellow, cyan lines) exhibit the same qualitative features
as those of SPEs in hBN previously reported in the literature (red and orange lines \cite{Tran2016a,Tran2016b}), namely intense
main bands together with weaker features red-shifted by $\sim 160$~meV. Importantly, spectra of both our samples and those
reported in the literature bear a striking resemblance to room-temperature fluorescence spectra recorded on single PAH
molecules, here shown for the representative PAH species terrylene (TRL, pink line \cite{Kulzer1999a}) and dibenzo-ovalene
(DBO, gray line \cite{Chen2019}).

We note that in Fig. 1m, we use designations for spectral features that differ from those commonly used in the field.
Specifically, previous work refers to the main and red-shifted peaks as ZPL and PSB, respectively, appropriate to the case of a
point-like defect interacting with the phonons of its crystalline host. Within the scenario of a molecular emission origin
demonstrated in this work, the main band derives primarily from the purely electronic ($0-0$) transition of an individual
molecule, while the red-shifted band originates from vibronic transitions of the molecule. We denote the energies of these
bands as $E_{0-0}$ and $E_{vibr}$, respectively, to reflect the underlying mechanism.

We present photophysical characterizations of two hBN samples that have undergone activation heat treatment after exposure to
anisole and ethanol in Supplementary Figures 1 and 2, respectively. PL intensity maps feature numerous bright, point-like
emission spots (Supplementary Fig. 1a and 2a). Many of these spots correspond to spectra comprising the typical multi-peak
structure with a peak spacing of $\sim 160$~meV (Supplementary Fig. 1b and 2b). The emission of single photons is demonstrated
by the observation of strong antibunching ($g^{(2)}(0) < 0.5$) in photon correlation measurements (Supplementary Fig. 1e and
2e,h,k). Fluorescence decay time constants cover the range of $2 - 12$ ns (Supplementary Fig. 1f and 2f,i), comparable to
lifetimes reported for hBN emitters in the literature
\cite{Tran2016a,Tran2016b,Tran2016c,Jungwirth2016a,Chejanovsky2016,Schell2016,Grosso2017,Exarhos2017}. We observe both
non-blinking and intermittent emitters (see also Supplementary Note 6).

We remark that the range of precursor substances that are ultimately converted to SPEs is not limited to preexisting
fluorescent impurities in the solvents used for sample immersion: high-temperature treatment creates emitters under hBN also if
spectroscopy-grade ethanol with very low levels of fluorescent impurities is used (see Supplementary Fig. 4). Further, we
observe emitter formation also for solvents other than anisole and ethanol (see Supplementary Fig. 5), consistent with the fact
that our proposed mechanism is insensitive to the specific chemical composition of organic precursors.

\subsection*{Emission centers are located at the hBN/substrate interface}

Next, we describe a series of experiments that demonstrate that the emitters are located at the hBN/substrate interface,
thereby proving that they do not result from point defects within the hBN crystal lattice. First we study the behavior of a
thick hBN sheet that contains suspended regions where hBN is not in contact with the substrate (schematic in Fig. 2a). Figure
2b shows a microscope photograph of this hBN flake (thickness range 260 - 300 nm), which was prepared with a single live cycle
involving immersion in anisole. High-resolution bright- and dark-field photographs (Fig. 2c,d) focus on a suspended hBN region
above a square marker etched into the \ce{SiO2} substrate (etch depth 50 nm). The PL intensity map of the same area (Fig. 2e)
shows a striking contrast between hBN areas supported by the substrate, where a high density of point-like emitters and uniform
background fluorescence are present, versus the absence of both in the suspended hBN area (red arrows in figure). Given the
substantial thickness of hBN ($\sim 260$ nm close to marker), this observation represents strong evidence that emitters are not
distributed homogeneously throughout the hBN bulk, but are instead located at or close to the hBN/\ce{SiO2} interface, in
accordance with our proposed mechanism of trapped PAH-like fluorescing molecules being the origin of PL emission. We note that
a great reduction of emitter density and background fluorescence in suspended thick hBN sheets has also been observed by
Exarhos \textit{et al.} \cite{Exarhos2017}, though the origin of this phenomenon was not explored in that work.

The ribbon-shaped non-fluorescing regions visible in Fig. 2e are ``intrinsically'' suspended areas of the hBN sheet, adjacent
to height steps that occur at the sheet's underside (see Supplementary Note 7). We use the same sample for an additional
demonstration that emission centers do not reside within the hBN bulk, showing that emitters persist through multiple
subsequent cycles of fluorine-plasma thinning of the hBN sheet (described in Supplementary Note 7).

In the following, we present an additional experiment that provides unambiguous evidence that emission centers are indeed
located strictly at the hBN/substrate interface. Specifically, we separate the hBN sheet from its substrate, and subsequently
investigate the sheet and the vacated substrate in isolation. Figure 3a illustrates our experimental procedure. After initial
characterization of an as-prepared hBN sheet that contains numerous emission centers, we deposit a 75 nm thick metal film (5 nm
chromium adhesion layer, 70 nm platinum) that acts as a mechanical support layer. Next, we attach double-sided adhesive tape to
the metal support film, lift the assembly of adhesive tape/metal film/hBN from the \ce{SiO2} substrate, flip it over, and
attach it to a glass slide. After completion of the procedure, the former underside of the hBN sheet is exposed and pointing
upward, and the hBN flake and the now-vacated substrate can be characterized separately.

The hBN sample discussed here was prepared with a single live cycle involving immersion in anisole. Figure 3b,c shows bright-
and darkfield images of the as-prepared, thick hBN flake (thickness 96 nm), which features a high density of point-like
emission centers in its PL intensity map (Fig. 3d), superimposed on a weaker fluorescence background. Subsequent to this
initial characterization, we detach the hBN flake from its substrate. Figure 3e,f present bright- and dark-field optical
photographs of the flipped-over hBN sheet on the metal support film (images mirrored horizontally for ease of comparison).
While mechanical strain exerted during the lift-up procedure leads to damage at the sheet edges and fine cracks within the
sheet, the overall shape of the hBN flake is preserved, and a clear one-to-one match of locations within the hBN sheet is
easily established.

Since hBN sheets do not exhibit significant color contrast with the metallic substrate in microscope images, we verify the
presence of hBN from an intensity map of its Raman spectrum (inset of Fig. 3g). Strikingly, the PL intensity map of the
flipped-over hBN sample (Fig. 3g), collected at the same location that previously featured a large number of emitters, is now
completely devoid of emission centers. Likewise, the fluorescence background has disappeared, and the only feature visible in
spectra recorded within the sheet is the Raman peak of hBN (Fig. 3l, red line).

Meanwhile, the now-vacated \ce{SiO2} substrate (Fig. 3h,i) has a PL intensity map (Fig. 3j) that exhibits significant
fluorescence intensity in exactly those areas that were previously covered by hBN. An AFM topography map of the same area (Fig.
3k) contains sharply delineated features that are several nanometers high and correspond to the edge of the removed hBN sheet,
presumably due to metal accumulated at the hBN sidewall during deposition, as well as more indistinct shapes reminiscent of a
dried liquid. However, the AFM scan does not detect any height differential across the former edge of the hBN sheet, which
corresponds to the PL boundary line in Fig. 3j, confirming the complete removal of hBN. PL spectra from the now-vacated
\ce{SiO2} substrate are enhanced but featureless, or show very broad maxima (Fig. 3l, blue and black lines), indicating that
the source of the uniform PL background of the as-prepared hBN sample (Fig. 3d) remains attached to the \ce{SiO2} substrate. We
note that the vacated \ce{SiO2} substrate exhibits neither bright, point-like spots nor sharp PL spectra suggestive of SPEs.
The absence of such features from both the lifted hBN sheet and the vacated substrate represents evidence that the previously
visible, bright emitters (Fig. 3d) are now exposed to ambient air, resulting in oxygen-induced photobleaching during laser
irradiation. (See Supplementary Note 8 for discussion of technical aspects and additional characterization.)

These characterizations, taken together with our demonstration that reducing an hBN sheet's thickness by a factor up to 25 by
plasma etching does not eliminate emitters proportionally (Supplementary Fig. 9), show conclusively that the emission centers
observed in the as-prepared sample were located not within the hBN volume, but trapped at the hBN/substrate interface. This
finding supports our hypothesis that emission centers are fluorescing molecules derived from organic residues during heat
treatment.

\subsection*{Universality of proposed SPE formation mechanism}

In the experiments described so far, we used thoroughly cleaned hBN samples, in which the appearance of SPEs was associated
with the deliberate exposure to organic solvents during sample processing. This mode of sample processing was chosen as a
convenient method for the controlled introduction of organic residues. However, PAH molecules can form from a large variety of
organic precursors at high temperature \cite{Norinaga2009,Viteri2019,Zhou2015}, and accordingly the mechanism of SPE formation
proposed in this work does not depend on the specific origin of organic material. Likewise, the mechanism is not exclusive to
hBN - rather, any 2D wide-bandgap insulator that can trap fluorescent molecules derived from organic residue and protect them
from oxygen-induced photo-bleaching during laser exposure is suitable, provided it is compatible with the sample preparation
procedure.

We first demonstrate that emitter formation in hBN can take place after exposure to organic materials other than solvents. For
mechanically exfoliated vdW crystals, glues from adhesive tape are a prominent source of organic residues if they are not
aggressively removed \cite{Garcia2012}. In Supplementary Fig. 6a-d, we show an exfoliated hBN sheet that underwent imperfect
cleavage, resulting in small amounts of adhesive residue trapped under the flake [white arrows in Supplementary Fig. 6a,b]. We
processed this sample according to a modified protocol that completely avoids exposure to organic solvents. Nevertheless, after
heat treatment at 850 \degC, its PL intensity map [Supplementary Fig. 6e] features numerous bright emission spots that
partially correlate with the location of adhesive residue in the original sheet. Spectra corresponding to emission centers
[Supplementary Fig. 6f] exhibit the same characteristic multi-peak structure, with a spacing $E_{0-0} - E_{vibr} \sim 160$ meV,
as those of emitters observed in solvent-exposed hBN samples, and of emitters reported in the literature. Evidently, the
presence of adhesive residue can facilitate the formation of emitters.

Next, we show that the class of SPEs emitting in the vicinity of 2 eV can form also in materials other than hBN. For this
demonstration, we select the layered insulator fluorophlogopite mica, with a bandgap $\sim 10.5$ eV.\cite{Buechner1975} The
mechanically exfoliated mica sheet shown in Fig. 4a was processed with the same solvent-free protocol as the hBN sample
described in the preceding paragraph. Subsequent to activation heat treatment at 850 \degC, its PL intensity map (Fig. 4b)
features bright, point-like emission spots, overlaying a weaker PL background. The two emitters highlighted in Fig. 4b exhibit
emission intensities of $2.7 \times 10^5$ and $4.0 \times 10^5$ photons/s (pink and yellow circles, respectively), comparable
to emitters associated with hBN sheets observed in this work and in the literature. The corresponding emission spectra (Fig.
4c) have a multi-peak structure, with a peak spacing $\sim 160$ meV, that is indistinguishable from that typically observed in
hBN. (See Supplementary Fig. 7 for additional data from this sample.)

In Fig. 4d-l, we show further characterizations of representative photostable emitters from the same mica sheet. The
observation of $g^{(2)}(0) < 0.5$ in photon correlation measurements (Fig. 4d,g,j) demonstrates the emission of single photons.
The fluorescence decay time constants of $7 - 9$ ns fall in the range that is reported for SPEs in hBN (Fig. 4e,h). The SPEs in
mica show typical saturation behavior, with $I \propto P/(P + P_{sat})$ (Fig. 4f), and a subset exhibits fluorescence
intermittency (Fig. 4i,l, see also Supplementary Movie 3).

In summary, we establish that neither the selection of hBN as host material, nor the specific identity of organic processing
residues are critical for the appearance of heterogeneous single photon emission $\sim 2$ eV.

\section*{Discussion}

Briefly summarizing, our proposal that SPEs associated with hBN that emit at variable energies $\sim 2$ eV are PAH-like
molecules of extrinsic origin is supported by the following experimental observations: (1) the appearance of emission centers
in hBN critically depends on the presence of organic residues during widely used activation heat treatment; (2) centers'
emission spectra closely resemble those of fluorescing PAH molecules; (3) emitters reside at the hBN/substrate interface; (4)
the same class of SPEs can also form below a separate wide-bandgap insulator host material, fluorophlogopite mica. Our proposal
can explain many of the anomalies of SPEs $\sim 2$ eV reported in the literature. Importantly, the large number of PAH species
that can form in an uncontrolled manner during high temperature treatment accounts in a natural fashion for the great
heterogeneity in reported emitter behavior, in particular the wide range in observed emission peak energies $E_{0-0}$ (referred
to as ``ZPL'' in most work).

The reported observations on SPEs $\sim 2$ eV in hBN directly mirror the properties of PAH molecules studied as SPEs, such as
fluorescence intermittency \cite{Kulzer1999a,Chen2019} and polarization behavior
\cite{Guttler1996,Pfab2004,Nicolet2007,Yuce2012}. Furthermore, the optical properties of single molecules are known to be
exquisitely sensitive to external perturbations, which corresponds to the marked response of single photon emission $\sim 2$ eV
from hBN to applied external fields that previous work has reported. Spectral shifts exhibited by SPEs in hBN upon the
application of external strain and pressure \cite{Grosso2017,Xue2018} are comparable to the shifts observed for single
pentacene and terrylene molecules in organic matrices in response to pressure \cite{Croci1993,Muller1995,Iwamoto1998}.
Application of an electric field leads to a Stark shift of the emission lines of SPEs in hBN, with emitter-specific linear and
quadratic Stark effects \cite{Noh2018,Mendelson2019,Nikolay2019a,Xia2019}. Such Stark effects also occur, with similar
coefficients, in single molecules of several PAH species studied as SPEs
\cite{Wild1992,Orrit1992,Kulzer1999b,Bauer2003,Moradi2019}.

Intriguingly, it has been shown that a small percentage of SPEs in hBN that emit at $\sim 2$ eV exhibit magnetic
field-dependent behavior \cite{Exarhos2019}. Optically detected magnetic resonance studies on isolated SPEs and ensembles
demonstrated emitter ground states with finite spin $S \geq 1/2$, with $g$ factors $\sim 2.0$.\cite{Mendelson2021,
Chejanovsky2021, Stern2022} These observations can be readily accommodated within the scenario of the emitters being PAH-like
molecules formed during high-temperature processing: theoretical and experimental work has explored a variety of polycyclic
hydrocarbon radicals with open-shell electron configurations that can possess ground states with $S \geq
1/2$.\cite{Gerson2003,Morita2011,Sun2012,Ahmed2022}. Such molecular radicals have $g$ factors close to the free electron value
$g_e = 2.0023$,\cite{Gerson2003} consistent with those observed for SPEs in hBN. Candidate open-shell molecules include the
well-studied series of triangular nanographene radicals \cite{Morita2011}. Optical spectra of phenalenyl ($S = 1/2$), the
smallest member of this series, exhibit a $0-0$ transition at $\sim 2.3$ eV in ensemble measurements \cite{Cofino1984,
OConnor2011}. However, since such radicals are short-lived in ambient environment due to their high chemical reactivity, the
synthesis of isolated single molecules, which has been demonstrated for the higher series members triangulene ($S=1$)
\cite{Pavlicek2017} and $\pi$-extended triangulene ($S=3/2$) \cite{Mishra2019}, has so far only been possible under ultrahigh
vacuum conditions, and no single molecule optical studies have been reported for these or other radicals. The creation of such
polycyclic hydrocarbon radicals under hBN sheets that provide inert encapsulation could thus represent a viable means for their
experimental exploration at the single molecule level.

It is worthwhile to reflect on the reasons why despite extensive studies of heterogeneous SPEs $\sim 2$ eV in hBN, their origin
in fluorescent organic molecules has not been recognized so far, and has even been discounted explicitly in some work
\cite{Chejanovsky2016}. We point to three distinct confounding factors: The ubiquity of organic residues in sample preparation,
the similar energy scales of hBN phonons and vibrational eigenmodes of PAH molecules, and the observation that even though hBN
defect sites may not be sources of emission $\sim 2$ eV, they can nevertheless lead to the appearance of SPEs by pinning PAH
molecules.

The first of these factors, organic processing residue, is present in virtually all reported sample preparation methods. The
most obvious case is liquid-exfoliated hBN, which has been used in many reports
\cite{Tran2016b,Jungwirth2016a,Shotan2016,Kozawa2023}. There, hBN sheets are suspended in an organic solvent (typically
ethanol), and sheet restacking is unavoidable during deposition onto a substrate, leading to trapped impurities both under and
between hBN sheets that cannot be removed. For sample preparation with mechanically exfoliated hBN, adhesive residues are
unavoidable. While some recipes use aggressive chemical or plasma treatment intended to remove organic impurities
\cite{Chejanovsky2016,Exarhos2017}, these methods do not affect residues that are deposited underneath or between hBN flakes
and are thus protected from external attack. This protection offered by hBN encapsulation also explains why emitters are
unaffected even by heat treatment in a reactive atmosphere \cite{Tran2016b}. Our work eliminates complications arising from
residues by combining a simple hBN sample geometry with an effective cleaning protocol. Critically, we choose mechanically
exfoliated hBN sheets that are free from folds and restacked sheets, such that adhesive residue is exposed on top surfaces and
can be easily removed. This enables us to establish residue-free hBN specimen as a baseline to which samples with deliberately
introduced organic contaminants can be contrasted.

Representing a separate confounding factor, the frequently observed additional PL spectral bands that are red-shifted by $\sim
160$~meV ($\approx 1300$~\invcm) from the main peak have commonly been interpreted as PSBs that reflect the interaction of
point defects with phonons of the hBN host \cite{Tran2016a,Chejanovsky2016,Jungwirth2016a,Khatri2019,Wigger2019}. Incidentally,
this energy spacing is also typical for vibrational eigenmodes of PAH molecules, clearly visible in their fluorescence spectra
\cite{Kulzer1999a,Chen2019} (see Fig. 1m). We reinterpret the red-shifted bands observed in hBN emitters as originating from
the vibronic transitions of trapped PAH molecules. This picture accounts both for the deviation of the spacing $E_{0-0} -
E_{vibr}$ from the phonon energy $E(\mathrm{E}_{2g}) = 1367$~\invcm\ of hBN, and for the slight spread of observed spacings,
expected if multiple PAH species coexist.

The final confounding factor is that even within the picture that PAH molecules are the source of SPEs, lattice defects in the
hBN crystal can nonetheless decisively influence their presence and distribution. In particular, the reported preference for
emitters to localize at the hBN sheet perimeter and extended structural defects
\cite{Tran2016c,Chejanovsky2016,Choi2016,Li2017,Xu2018} can be explained by the higher binding energy of organic molecules at
under-coordinated defect sites of hBN.

In Supplementary Note 9, we discuss the observations on SPEs $\sim 2$ eV reported in previous work for a range of sample
preparation methods, including emitters in epitaxially grown hBN, as well as SPEs produced by electron irradiation, and by
strain. Our proposal that fluorescent molecules are the origin of heterogeneous SPEs $\sim 2$ eV appears capable of
accommodating these diverse results. As an intriguing exception, we highlight recent reports of \textit{homogeneous}, electron
beam-induced SPEs in hBN that are characterized by invariant emission maxima at 575 nm, and by polarization axes that exhibit
fixed relationships with the hBN crystal axes \cite{Kumar2022,Kumar2023}. These observations are more consistent with single
photon emission from hBN lattice defects, thus suggesting that multiple distinct SPE formation mechanisms seem to coexist in
the energy range $\sim 2$ eV.

\section*{Conclusions}

We have presented experimental evidence that unambiguously links the appearance of heterogeneous emission centers $\sim 2$ eV
in hBN to the presence of organic residues, and that strongly supports the scenario that single-photon emission in this energy
range originates from extrinsically formed fluorescent molecules. This picture differs significantly from models explored in
previous work that center around the notion that microscopic hBN lattice defects are the source of single-photon emission.
While we have to leave open the possibility that hBN crystal defects may contribute to the observed heterogeneity, we remark
that our model appears able to account comprehensively for a large number of disparate observations on the heterogeneous SPEs
$\sim 2$ eV reported in the literature. Our specific proposal that PAH-like molecules are the emitting entities provides clear
explanations for the important role that heat treatment is known to play in emitter formation, for the characteristic spectral
shapes of SPEs observed in hBN, and for the thus far unexplained great heterogeneity in the photophysical properties of
emitters.

Our result implies a reassessment of the role of hBN, in the sense that it is not necessarily required as an active and
indispensable host for the most widely studied, heterogeneous SPEs $\sim 2$ eV. Nonetheless, we believe that hBN will retain
its important role also for this emitter class, due to its advantageous properties as an atomically flat, thin, inert
encapsulation that protects emitters from oxygen-induced photo-bleaching, as demonstrated in a wealth of previous work.
Moreover, hBN may even represent a viable alternative to the organic host crystals typically used when studying single photon
emission from individual PAH molecules \cite{Toninelli2021}. Most importantly, the change of perspective introduced by our work
will accelerate the development of hBN-based quantum photonic technologies, by guiding a shift from uncontrolled emitter
generation towards rational selection of fluorophore species and their controlled placement, as demanded by many applications
in emerging quantum technologies.

\section*{Methods}

\subsection*{Sample preparation}

As substrates, we use silicon wafer pieces capped with 300 nm dry oxide. Etched orientation markers in the substrate are
patterned by photolithography, followed by an \ce{SF6} plasma etch, with an etch depth of 50 nm. Prior to hBN exfoliation, we
ensure that substrates are free from photoresist and ambient residues by sequential immersion in acetone and trichloroethylene,
followed by baking in oxygen atmosphere at 700 \degC. We prepare hBN flakes by mechanical cleavage of bulk hBN single crystals
grown by a high-temperature high-pressure process \cite{Watanabe2004,Taniguchi2007}, using Ultron 1007R wafer dicing tape for
exfoliation and deposition on substrates, followed by immersion in trichloroethylene which greatly reduces adhesive residues of
the dicing tape used \cite{Kuo2016}.

We use two complementary sample treatment protocols to study the role of organic residues. In a control cycle, samples first
undergo a cleaning step (heat treatment in argon/10\%(mol) oxygen atmosphere at 700 \degC, atmospheric pressure, for $\sim 4$
hours), followed after cool-down by an `activation' step (850 \degC\ in $\sim 1$~Torr argon (5N), for $\sim 2$ hours) that has
been widely used in previous work to produce SPEs. In a `live' cycle, we perform the same steps, however with the additional
step of immersion in organic solvent inserted between oxidative cleaning and activation. Diffusion of the solvent between hBN
sheets and the \ce{SiO2} substrate, which we demonstrated in previous work \cite{Gasparutti2020}, enables the controlled
introduction of organic residue at the hBN/substrate interface. It has been shown that the cleaning step does not lead to
oxidative damage of hBN flakes \cite{Li2014}.

Solvents used for immersion are ethanol, anisole, dimethylformamide (DMF), and N-methyl pyrrolidone (NMP). To avoid
uncontrolled contamination due to leaching when performing sample immersion, we pour solvents from their original glass bottles
into borosilicate glassware (which is used exclusively for this operation) directly, without the use of syringes or pipets, and
handle samples with stainless steel tweezers. We acquire optical microscope photographs of the studied hBN flakes both before
and after immersion, to detect instances of flake detachment and movement. Flake detachment during immersion is common for DMF
and NMP, but not for anisole and ethanol (see Supplementary Note 4 for discussion). Flakes that detached are not studied
further.

The following solvents studied as starting materials for emitter generation were purchased from Daejung Chemicals, Korea (with
product number, grade, purity): ethanol (P/N 4023-4100, anhydrous, guaranteed reagent, assay $\geq 99.98$\%), anisole (P/N
1089-4400, extra pure, assay $99.9$\%), DMF (P/N 3057-7100, special guaranteed reagent, assay $\geq 99.9$\%), NMP (P/N
5575-4100, guaranteed reagent, assay $99.9$\%). Spectroscopy-grade ethanol was bought from Merck: Uvasol ethanol (P/N 100980,
assay $\geq 99.9$\%). Solvents used during cleaning steps were purchased from Daejung Chemicals: trichloroethylene (P/N
8553-4400, extra pure, assay $99.2$\%) and acetone (P/N 1009-2304, HPLC solvent, assay $\geq 99.9$\%). Process gases are
procured from JS Gas, Korea. All furnace treatment takes place within a fused silica tube that is used exclusively for hBN
samples to prevent contamination. During activation treatment, the furnace is pumped by a dry scroll pump.

Synthetic fluorophlogopite mica (\ce{KMg3(AlSi3)O10F2}) was manufactured by Taiyuan Fluorphlogopite Mica Co. Ltd. (plate
dimensions $10 \times 10 \times 0.2$ mm). Mechanical exfoliation of mica is performed identically to that of hBN, albeit
without subsequent immersion in trichloroethylene. Since almost no diffusion of organic solvent takes place along the interface
between very hydrophilic mica and \ce{SiO2}, exfoliated mica samples are processed using a solvent-free protocol that comprises
a cleaning step (heat treatment in argon/10\%(mol) oxygen atmosphere at 700 \degC, atmospheric pressure) followed by an
`activation' step (850 \degC\ in $\sim 1$~Torr argon (5N)). Emitters formed in mica samples (Fig. 4, Supplementary Fig. 7), and
in identically prepared hBN control samples (Supplementary Fig. 6), reflect the presence of adhesive residues under imperfectly
cleaved sheets.

\subsection*{Sample characterization}

Photoluminescence (PL) data are collected on a WITec alpha-300 microscope, using a 100x objective (NA = 0.9) and a cw-laser of
wavelength 532~nm. PL areal scans are acquired with an incident laser power 480~$\mu$W, except where stated otherwise, and
exposure time 100 ms/pixel. PL intensity maps are calculated by integrating the spectral range $1.88 - 2.19$ eV, excluding the
narrow hBN Raman peak at 2.16 eV. During prolonged PL spectra acquisition for the study of fluorescence intermittency, we use a
laser power 35~$\mu$W. All PL spectra are corrected for the wavelength-dependent reflection efficiency of the spectrometer
grating (600 grooves/mm, 500 nm blazing) \cite{Hollrichter2010} and quantum efficiency of the CCD detector (Andor DV401-BV).

Wide-field fluorescence movies are acquired under continuous cw-laser condition at 532 nm wavelength, using a Coherent
Sapphire-532-200 laser, and processed with the ImageJ/Fiji software package \cite{Schindelin2012}. We perform photon
correlation, time-resolved fluorescence, and saturation measurements using a PicoBlade 532 laser of wavelength 532 nm, with a
repetition rate of 8.2 MHz and a pulse width of 10 ps. Photon correlation data are acquired using a HydraHarp 400 photon
counting module, and are analyzed using the fit function
$$
I(t) = \sum_N \alpha_N \exp \left( \frac{- |t - N \times t_{offs}| }{ \tau_N } \right)
$$
with $t_{offs} = 1/\mathrm{(8.2 \, MHz)}$. Numerical values of $g^{(2)}(0)$ are evaluated by integration of photon correlation
counts. Integration of the fit function leads to identical results. Additional photon correlation measurements are collected
using the cw-laser. These data are normalized at large delay times, and analyzed with the fit function $I(t) = 1 - a \exp(
-|t|/\tau )$. Background-subtracted saturation data follow the relationship $I(P) = I_0 \times P/(P+P_{sat})$. Time-resolved
fluorescence data are analyzed with mono- and biexponential fit functions.

For AFM characterization, we use a Park Systems XE-7 AFM, with silicon cantilevers in intermittent-contact mode. PL intensity
and AFM topography maps were processed using the Gwyddion software package \cite{Necas2012}. All PL spectra presented
correspond to single pixels in the accompanying intensity maps. Raman spectra (Fig. 3m, Supplementary Fig. 5) are obtained by
averaging $7 \times 7$ spectra corresponding to neighboring pixels to improve signal/noise ratio of the $\mathrm{E}_{2g}$ Raman
peak of hBN, and to better resolve the D- and G-bands of graphitic layers (Supplementary Fig. 5). Raman spectra shown in
Supplementary Fig. 11 are calculated by averaging $> 8,000$ spectra each. Optical bright- and dark-field microscope images are
collected on a Leica DM2700 M with LED white light illumination.

\subsection*{Metal-based hBN pickup and transfer}

The continuous metal support layer that is used to pick up hBN sheets is composed of a 5 nm thick chromium adhesion layer and
70 nm platinum. The metal film is deposited onto the samples using an electron beam evaporator. We attach a double-sided
adhesive strip (3M Scotch double-sided tape) onto the support layer, lift up the assembly such that the metal film and hBN
flakes detach from the substrate, and attach the assembly upside-down to a glass slide as a rigid support.

\section*{Supporting information available}

The Supporting Information PDF file describes additional experimental methods and contains Supplementary Notes: empirical
observations that implicate organic fluorophores as SPE source, domain of applicability of proposed emission mechanism,
characterization of single photon emitters, exploring the role of solvents in emitter creation, emitters in hBN and mica
derived from adhesive residue, emitter intermittency and spectral diffusion, plasma-thinning of hBN does not eliminate
emitters, proving emitter location at hBN/substrate interface by sample disassembly, additional discussion of previous work.

Supplementary Movies S1-S3 show widefield fluorescence movies of emitters observed in hBN sheets exposed to the organic
solvents anisole and ethanol during sample preparation, and of emitters in fluoro\-phlogopite mica.

\section*{Author contributions}

M.N. conceived the project. M.N. and X.W. fabricated and characterized samples and analyzed data. L.M.-I. and S.G. performed
SPE characterizations. K.W. and T.T. synthesized single crystal hBN. M.N., S.G. and Y.H.L. directed research. All authors were
involved in discussions. M.N. wrote the manuscript, with contributions from all authors.

\section*{Competing interests}

The authors declare no competing interests.

\section*{Acknowledgments}

This work was supported by the Institute for Basic Science (grant IBS-R011-D1) and the Max Planck Society. K.W. and T.T.
acknowledge support from the Elemental Strategy Initiative conducted by the MEXT, Japan, A3 Foresight by JSPS and JSPS KAKENHI
(grant numbers JP19H05790 and JP20H00354). L.M.-I. and S.G. thank V. Sandoghdar for continuous support.

\bibliography{hBN-emitters-arxiv}

\providecommand{\latin}[1]{#1}
\makeatletter
\providecommand{\doi}
  {\begingroup\let\do\@makeother\dospecials
  \catcode`\{=1 \catcode`\}=2 \doi@aux}
\providecommand{\doi@aux}[1]{\endgroup\texttt{#1}}
\makeatother
\providecommand*\mcitethebibliography{\thebibliography}
\csname @ifundefined\endcsname{endmcitethebibliography}
  {\let\endmcitethebibliography\endthebibliography}{}
\begin{mcitethebibliography}{86}
\providecommand*\natexlab[1]{#1}
\providecommand*\mciteSetBstSublistMode[1]{}
\providecommand*\mciteSetBstMaxWidthForm[2]{}
\providecommand*\mciteBstWouldAddEndPuncttrue
  {\def\EndOfBibitem{\unskip.}}
\providecommand*\mciteBstWouldAddEndPunctfalse
  {\let\EndOfBibitem\relax}
\providecommand*\mciteSetBstMidEndSepPunct[3]{}
\providecommand*\mciteSetBstSublistLabelBeginEnd[3]{}
\providecommand*\EndOfBibitem{}
\mciteSetBstSublistMode{f}
\mciteSetBstMaxWidthForm{subitem}{(\alph{mcitesubitemcount})}
\mciteSetBstSublistLabelBeginEnd
  {\mcitemaxwidthsubitemform\space}
  {\relax}
  {\relax}

\bibitem[Zhang \latin{et~al.}(2020)Zhang, Cheng, Chou, and Gali]{Zhang2020a}
Zhang,~G.; Cheng,~Y.; Chou,~J.-P.; Gali,~A. {Material Platforms for Defect
  Qubits and Single-Photon Emitters}. \emph{Appl. Phys. Rev} \textbf{2020},
  \emph{7}, 031308\relax
\mciteBstWouldAddEndPuncttrue
\mciteSetBstMidEndSepPunct{\mcitedefaultmidpunct}
{\mcitedefaultendpunct}{\mcitedefaultseppunct}\relax
\EndOfBibitem
\bibitem[Tran \latin{et~al.}(2016)Tran, Bray, Ford, Toth, and
  Aharonovich]{Tran2016a}
Tran,~T.~T.; Bray,~K.; Ford,~M.~J.; Toth,~M.; Aharonovich,~I. {Quantum Emission
  from Hexagonal Boron Nitride Monolayers}. \emph{Nat. Nanotechnol.}
  \textbf{2016}, \emph{11}, 37\relax
\mciteBstWouldAddEndPuncttrue
\mciteSetBstMidEndSepPunct{\mcitedefaultmidpunct}
{\mcitedefaultendpunct}{\mcitedefaultseppunct}\relax
\EndOfBibitem
\bibitem[Toth and Aharonovich(2019)Toth, and Aharonovich]{Toth2019}
Toth,~M.; Aharonovich,~I. {Single Photon Sources in Atomically Thin Materials}.
  \emph{Annu. Rev. Phys. Chem.} \textbf{2019}, \emph{70}, 123--142\relax
\mciteBstWouldAddEndPuncttrue
\mciteSetBstMidEndSepPunct{\mcitedefaultmidpunct}
{\mcitedefaultendpunct}{\mcitedefaultseppunct}\relax
\EndOfBibitem
\bibitem[Caldwell \latin{et~al.}(2019)Caldwell, Aharonovich, Cassabois, Edgar,
  Gil, and Basov]{Caldwell2019}
Caldwell,~J.~D.; Aharonovich,~I.; Cassabois,~G.; Edgar,~J.~H.; Gil,~B.;
  Basov,~D. {Photonics with Hexagonal Boron Nitride}. \emph{Nat. Rev. Mater.}
  \textbf{2019}, \emph{4}, 552--567\relax
\mciteBstWouldAddEndPuncttrue
\mciteSetBstMidEndSepPunct{\mcitedefaultmidpunct}
{\mcitedefaultendpunct}{\mcitedefaultseppunct}\relax
\EndOfBibitem
\bibitem[Sajid \latin{et~al.}(2020)Sajid, Ford, and Reimers]{Sajid2020a}
Sajid,~A.; Ford,~M.~J.; Reimers,~J.~R. {Single-Photon Emitters in Hexagonal
  Boron Nitride: a Review of Progress}. \emph{Rep. Prog. Phys.} \textbf{2020},
  \emph{83}, 044501\relax
\mciteBstWouldAddEndPuncttrue
\mciteSetBstMidEndSepPunct{\mcitedefaultmidpunct}
{\mcitedefaultendpunct}{\mcitedefaultseppunct}\relax
\EndOfBibitem
\bibitem[Tran \latin{et~al.}(2016)Tran, Elbadawi, Totonjian, Lobo, Grosso,
  Moon, Englund, Ford, Aharonovich, and Toth]{Tran2016b}
Tran,~T.~T.; Elbadawi,~C.; Totonjian,~D.; Lobo,~C.~J.; Grosso,~G.; Moon,~H.;
  Englund,~D.~R.; Ford,~M.~J.; Aharonovich,~I.; Toth,~M. {Robust Multicolor
  Single Photon Emission from Point Defects in Hexagonal Boron Nitride}.
  \emph{ACS Nano} \textbf{2016}, \emph{10}, 7331--7338\relax
\mciteBstWouldAddEndPuncttrue
\mciteSetBstMidEndSepPunct{\mcitedefaultmidpunct}
{\mcitedefaultendpunct}{\mcitedefaultseppunct}\relax
\EndOfBibitem
\bibitem[Jungwirth \latin{et~al.}(2016)Jungwirth, Calderon, Ji, Spencer,
  Flatt{\'e}, and Fuchs]{Jungwirth2016a}
Jungwirth,~N.~R.; Calderon,~B.; Ji,~Y.; Spencer,~M.~G.; Flatt{\'e},~M.~E.;
  Fuchs,~G.~D. {Temperature Dependence of Wavelength Selectable Zero-Phonon
  Emission from Single Defects in Hexagonal Boron Nitride}. \emph{Nano Lett.}
  \textbf{2016}, \emph{16}, 6052--6057\relax
\mciteBstWouldAddEndPuncttrue
\mciteSetBstMidEndSepPunct{\mcitedefaultmidpunct}
{\mcitedefaultendpunct}{\mcitedefaultseppunct}\relax
\EndOfBibitem
\bibitem[Shotan \latin{et~al.}(2016)Shotan, Jayakumar, Considine, Mackoit,
  Fedder, Wrachtrup, Alkauskas, Doherty, Menon, and Meriles]{Shotan2016}
Shotan,~Z.; Jayakumar,~H.; Considine,~C.~R.; Mackoit,~M.; Fedder,~H.;
  Wrachtrup,~J.; Alkauskas,~A.; Doherty,~M.~W.; Menon,~V.~M.; Meriles,~C.~A.
  {Photoinduced Modification of Single-Photon Emitters in Hexagonal Boron
  Nitride}. \emph{ACS Photonics} \textbf{2016}, \emph{3}, 2490--2496\relax
\mciteBstWouldAddEndPuncttrue
\mciteSetBstMidEndSepPunct{\mcitedefaultmidpunct}
{\mcitedefaultendpunct}{\mcitedefaultseppunct}\relax
\EndOfBibitem
\bibitem[Kozawa \latin{et~al.}(2023)Kozawa, Li, Ichihara, Rajan, Gong, He,
  Koman, Zeng, Kuehne, Silmore, Parviz, Liu, Liu, Faucher, Yuan, Warner,
  Blankschtein, and Strano]{Kozawa2023}
Kozawa,~D.; Li,~S.~X.; Ichihara,~T.; Rajan,~A.~G.; Gong,~X.; He,~G.;
  Koman,~V.~B.; Zeng,~Y.; Kuehne,~M.; Silmore,~K.~S.; Parviz,~D.; Liu,~P.;
  Liu,~A.~T.; Faucher,~S.; Yuan,~Z.; Warner,~J.; Blankschtein,~D.;
  Strano,~M.~S. {Discretized Hexagonal Boron Nitride Quantum Emitters and Their
  Chemical Interconversion}. \emph{Nanotechnology} \textbf{2023}, \emph{34},
  115702\relax
\mciteBstWouldAddEndPuncttrue
\mciteSetBstMidEndSepPunct{\mcitedefaultmidpunct}
{\mcitedefaultendpunct}{\mcitedefaultseppunct}\relax
\EndOfBibitem
\bibitem[Tran \latin{et~al.}(2016)Tran, Zachreson, Berhane, Bray, Sandstrom,
  Li, Taniguchi, Watanabe, Aharonovich, and Toth]{Tran2016c}
Tran,~T.~T.; Zachreson,~C.; Berhane,~A.~M.; Bray,~K.; Sandstrom,~R.~G.;
  Li,~L.~H.; Taniguchi,~T.; Watanabe,~K.; Aharonovich,~I.; Toth,~M. {Quantum
  Emission from Defects in Single-Crystalline Hexagonal Boron Nitride}.
  \emph{Phys. Rev. Appl.} \textbf{2016}, \emph{5}, 034005\relax
\mciteBstWouldAddEndPuncttrue
\mciteSetBstMidEndSepPunct{\mcitedefaultmidpunct}
{\mcitedefaultendpunct}{\mcitedefaultseppunct}\relax
\EndOfBibitem
\bibitem[Chejanovsky \latin{et~al.}(2016)Chejanovsky, Rezai, Paolucci, Kim,
  Rendler, Rouabeh, F{\'a}varo~de Oliveira, Herlinger, Denisenko, Yang,
  Gerhardt, Finkler, Smet, and Wrachtrup]{Chejanovsky2016}
Chejanovsky,~N.; Rezai,~M.; Paolucci,~F.; Kim,~Y.; Rendler,~T.; Rouabeh,~W.;
  F{\'a}varo~de Oliveira,~F.; Herlinger,~P.; Denisenko,~A.; Yang,~S.;
  Gerhardt,~I.; Finkler,~A.; Smet,~J.~H.; Wrachtrup,~J. {Structural Attributes
  and Photodynamics of Visible Spectrum Quantum Emitters in Hexagonal Boron
  Nitride}. \emph{Nano Lett.} \textbf{2016}, \emph{16}, 7037--7045\relax
\mciteBstWouldAddEndPuncttrue
\mciteSetBstMidEndSepPunct{\mcitedefaultmidpunct}
{\mcitedefaultendpunct}{\mcitedefaultseppunct}\relax
\EndOfBibitem
\bibitem[Choi \latin{et~al.}(2016)Choi, Tran, Elbadawi, Lobo, Wang, Juodkazis,
  Seniutinas, Toth, and Aharonovich]{Choi2016}
Choi,~S.; Tran,~T.~T.; Elbadawi,~C.; Lobo,~C.; Wang,~X.; Juodkazis,~S.;
  Seniutinas,~G.; Toth,~M.; Aharonovich,~I. {Engineering and Localization of
  Quantum Emitters in Large Hexagonal Boron Nitride Layers}. \emph{ACS Appl.
  Mater. Interfaces} \textbf{2016}, \emph{8}, 29642--29648\relax
\mciteBstWouldAddEndPuncttrue
\mciteSetBstMidEndSepPunct{\mcitedefaultmidpunct}
{\mcitedefaultendpunct}{\mcitedefaultseppunct}\relax
\EndOfBibitem
\bibitem[Xu \latin{et~al.}(2018)Xu, Elbadawi, Tran, Kianinia, Li, Liu, Hoffman,
  Nguyen, Kim, Edgar, Wu, Song, Ali, Ford, Toth, and Aharonovich]{Xu2018}
Xu,~Z.-Q.; Elbadawi,~C.; Tran,~T.~T.; Kianinia,~M.; Li,~X.; Liu,~D.;
  Hoffman,~T.~B.; Nguyen,~M.; Kim,~S.; Edgar,~J.~H.; Wu,~X.; Song,~L.; Ali,~S.;
  Ford,~M.; Toth,~M.; Aharonovich,~I. {Single Photon Emission from Plasma
  Treated 2D Hexagonal Boron Nitride}. \emph{Nanoscale} \textbf{2018},
  \emph{10}, 7957--7965\relax
\mciteBstWouldAddEndPuncttrue
\mciteSetBstMidEndSepPunct{\mcitedefaultmidpunct}
{\mcitedefaultendpunct}{\mcitedefaultseppunct}\relax
\EndOfBibitem
\bibitem[Fischer \latin{et~al.}(2021)Fischer, Caridad, Sajid, Ghaderzadeh,
  Ghorbani-Asl, Gammelgaard, B{\o}ggild, Thygesen, Krasheninnikov, Xiao, Wubs,
  and Stenger]{Fischer2021}
Fischer,~M.; Caridad,~J.~M.; Sajid,~A.; Ghaderzadeh,~S.; Ghorbani-Asl,~M.;
  Gammelgaard,~L.; B{\o}ggild,~P.; Thygesen,~K.~S.; Krasheninnikov,~A.~V.;
  Xiao,~S.; Wubs,~M.; Stenger,~N. Controlled Generation of Luminescent Centers
  in Hexagonal Boron Nitride by Irradiation Engineering. \emph{Sci. Adv.}
  \textbf{2021}, \emph{7}, eabe7138\relax
\mciteBstWouldAddEndPuncttrue
\mciteSetBstMidEndSepPunct{\mcitedefaultmidpunct}
{\mcitedefaultendpunct}{\mcitedefaultseppunct}\relax
\EndOfBibitem
\bibitem[Li \latin{et~al.}(2017)Li, Shepard, Cupo, Camporeale, Shayan, Luo,
  Meunier, and Strauf]{Li2017}
Li,~X.; Shepard,~G.~D.; Cupo,~A.; Camporeale,~N.; Shayan,~K.; Luo,~Y.;
  Meunier,~V.; Strauf,~S. {Nonmagnetic Quantum Emitters in Boron Nitride with
  Ultranarrow and Sideband-Free Emission Spectra}. \emph{ACS Nano}
  \textbf{2017}, \emph{11}, 6652--6660\relax
\mciteBstWouldAddEndPuncttrue
\mciteSetBstMidEndSepPunct{\mcitedefaultmidpunct}
{\mcitedefaultendpunct}{\mcitedefaultseppunct}\relax
\EndOfBibitem
\bibitem[Stern \latin{et~al.}(2019)Stern, Wang, Fan, Mizuta, Stewart, Needham,
  Roberts, Wai, Ginsberg, Klenerman, Hofmann, and Lee]{Stern2019}
Stern,~H.~L.; Wang,~R.; Fan,~Y.; Mizuta,~R.; Stewart,~J.~C.; Needham,~L.-M.;
  Roberts,~T.~D.; Wai,~R.; Ginsberg,~N.~S.; Klenerman,~D.; Hofmann,~S.;
  Lee,~S.~F. {Spectrally Resolved Photodynamics of Individual Emitters in
  Large-Area Monolayers of Hexagonal Boron Nitride}. \emph{ACS Nano}
  \textbf{2019}, \emph{13}, 4538--4547\relax
\mciteBstWouldAddEndPuncttrue
\mciteSetBstMidEndSepPunct{\mcitedefaultmidpunct}
{\mcitedefaultendpunct}{\mcitedefaultseppunct}\relax
\EndOfBibitem
\bibitem[Mendelson \latin{et~al.}(2019)Mendelson, Xu, Tran, Kianinia, Scott,
  Bradac, Aharonovich, and Toth]{Mendelson2019}
Mendelson,~N.; Xu,~Z.-Q.; Tran,~T.~T.; Kianinia,~M.; Scott,~J.; Bradac,~C.;
  Aharonovich,~I.; Toth,~M. {Engineering and Tuning of Quantum Emitters in
  Few-Layer Hexagonal Boron Nitride}. \emph{ACS Nano} \textbf{2019}, \emph{13},
  3132--3140\relax
\mciteBstWouldAddEndPuncttrue
\mciteSetBstMidEndSepPunct{\mcitedefaultmidpunct}
{\mcitedefaultendpunct}{\mcitedefaultseppunct}\relax
\EndOfBibitem
\bibitem[Mendelson \latin{et~al.}(2021)Mendelson, Morales-Inostroza, Li,
  Ritika, Nguyen, Loyola-Echeverria, Kim, G{\"o}tzinger, Toth, and
  Aharonovich]{Mendelson2020}
Mendelson,~N.; Morales-Inostroza,~L.; Li,~C.; Ritika,~R.; Nguyen,~M. A.~P.;
  Loyola-Echeverria,~J.; Kim,~S.; G{\"o}tzinger,~S.; Toth,~M.; Aharonovich,~I.
  {Grain Dependent Growth of Bright Quantum Emitters in Hexagonal Boron
  Nitride}. \emph{Adv. Opt. Mater.} \textbf{2021}, \emph{9}, 2001271\relax
\mciteBstWouldAddEndPuncttrue
\mciteSetBstMidEndSepPunct{\mcitedefaultmidpunct}
{\mcitedefaultendpunct}{\mcitedefaultseppunct}\relax
\EndOfBibitem
\bibitem[Li \latin{et~al.}(2021)Li, Mendelson, Ritika, Chen, Xu, Toth, and
  Aharonovich]{Li2021}
Li,~C.; Mendelson,~N.; Ritika,~R.; Chen,~Y.; Xu,~Z.-Q.; Toth,~M.;
  Aharonovich,~I. {Scalable and Deterministic Fabrication of Quantum Emitter
  Arrays from Hexagonal Boron Nitride}. \emph{Nano Lett.} \textbf{2021},
  \emph{21}, 3626--3632\relax
\mciteBstWouldAddEndPuncttrue
\mciteSetBstMidEndSepPunct{\mcitedefaultmidpunct}
{\mcitedefaultendpunct}{\mcitedefaultseppunct}\relax
\EndOfBibitem
\bibitem[Grosso \latin{et~al.}(2017)Grosso, Moon, Lienhard, Ali, Efetov,
  Furchi, Jarillo-Herrero, Ford, Aharonovich, and Englund]{Grosso2017}
Grosso,~G.; Moon,~H.; Lienhard,~B.; Ali,~S.; Efetov,~D.~K.; Furchi,~M.~M.;
  Jarillo-Herrero,~P.; Ford,~M.~J.; Aharonovich,~I.; Englund,~D. {Tunable and
  High-Purity Room Temperature Single-Photon Emission from Atomic Defects in
  Hexagonal Boron Nitride}. \emph{Nat. Commun.} \textbf{2017}, \emph{8},
  705\relax
\mciteBstWouldAddEndPuncttrue
\mciteSetBstMidEndSepPunct{\mcitedefaultmidpunct}
{\mcitedefaultendpunct}{\mcitedefaultseppunct}\relax
\EndOfBibitem
\bibitem[Mendelson \latin{et~al.}(2021)Mendelson, Chugh, Reimers, Cheng,
  Gottscholl, Long, Mellor, Zettl, Dyakonov, Beton, Novikov, Jagadish, Tan,
  Ford, Toth, Bradac, and Aharonovich]{Mendelson2021}
Mendelson,~N.; Chugh,~D.; Reimers,~J.~R.; Cheng,~T.~S.; Gottscholl,~A.;
  Long,~H.; Mellor,~C.~J.; Zettl,~A.; Dyakonov,~V.; Beton,~P.~H.;
  Novikov,~S.~V.; Jagadish,~C.; Tan,~H.~H.; Ford,~M.~J.; Toth,~M.; Bradac,~C.;
  Aharonovich,~I. {Identifying Carbon as the Source of Visible Single-Photon
  Emission from Hexagonal Boron Nitride}. \emph{Nat. Mater.} \textbf{2021},
  \emph{20}, 321--328\relax
\mciteBstWouldAddEndPuncttrue
\mciteSetBstMidEndSepPunct{\mcitedefaultmidpunct}
{\mcitedefaultendpunct}{\mcitedefaultseppunct}\relax
\EndOfBibitem
\bibitem[Proscia \latin{et~al.}(2018)Proscia, Shotan, Jayakumar, Reddy, Cohen,
  Dollar, Alkauskas, Doherty, Meriles, and Menon]{Proscia2018}
Proscia,~N.~V.; Shotan,~Z.; Jayakumar,~H.; Reddy,~P.; Cohen,~C.; Dollar,~M.;
  Alkauskas,~A.; Doherty,~M.; Meriles,~C.~A.; Menon,~V.~M. {Near-Deterministic
  Activation of Room-Temperature Quantum Emitters in Hexagonal Boron Nitride}.
  \emph{Optica} \textbf{2018}, \emph{5}, 1128--1134\relax
\mciteBstWouldAddEndPuncttrue
\mciteSetBstMidEndSepPunct{\mcitedefaultmidpunct}
{\mcitedefaultendpunct}{\mcitedefaultseppunct}\relax
\EndOfBibitem
\bibitem[Mart{\'i}nez \latin{et~al.}(2016)Mart{\'i}nez, Pelini, Waselowski,
  Maze, Gil, Cassabois, and Jacques]{Martinez2016}
Mart{\'i}nez,~L.~J.; Pelini,~T.; Waselowski,~V.; Maze,~J.~R.; Gil,~B.;
  Cassabois,~G.; Jacques,~V. {Efficient Single Photon Emission from a
  High-Purity Hexagonal Boron Nitride Crystal}. \emph{Phys. Rev. B}
  \textbf{2016}, \emph{94}, 121405\relax
\mciteBstWouldAddEndPuncttrue
\mciteSetBstMidEndSepPunct{\mcitedefaultmidpunct}
{\mcitedefaultendpunct}{\mcitedefaultseppunct}\relax
\EndOfBibitem
\bibitem[Exarhos \latin{et~al.}(2017)Exarhos, Hopper, Grote, Alkauskas, and
  Bassett]{Exarhos2017}
Exarhos,~A.~L.; Hopper,~D.~A.; Grote,~R.~R.; Alkauskas,~A.; Bassett,~L.~C.
  {Optical Signatures of Quantum Emitters in Suspended Hexagonal Boron
  Nitride}. \emph{ACS Nano} \textbf{2017}, \emph{11}, 3328--3336\relax
\mciteBstWouldAddEndPuncttrue
\mciteSetBstMidEndSepPunct{\mcitedefaultmidpunct}
{\mcitedefaultendpunct}{\mcitedefaultseppunct}\relax
\EndOfBibitem
\bibitem[Schell \latin{et~al.}(2016)Schell, Tran, Takashima, Takeuchi, and
  Aharonovich]{Schell2016}
Schell,~A.~W.; Tran,~T.~T.; Takashima,~H.; Takeuchi,~S.; Aharonovich,~I.
  {Non-Linear Excitation of Quantum Emitters in Hexagonal Boron Nitride
  Multilayers}. \emph{APL Photonics} \textbf{2016}, \emph{1}, 091302\relax
\mciteBstWouldAddEndPuncttrue
\mciteSetBstMidEndSepPunct{\mcitedefaultmidpunct}
{\mcitedefaultendpunct}{\mcitedefaultseppunct}\relax
\EndOfBibitem
\bibitem[Xue \latin{et~al.}(2018)Xue, Wang, Tan, Zhang, Yu, Ding, Jiang, Dou,
  Shi, and Sun]{Xue2018}
Xue,~Y.; Wang,~H.; Tan,~Q.; Zhang,~J.; Yu,~T.; Ding,~K.; Jiang,~D.; Dou,~X.;
  Shi,~J.-j.; Sun,~B.-q. {Anomalous Pressure Characteristics of Defects in
  Hexagonal Boron Nitride Flakes}. \emph{ACS Nano} \textbf{2018}, \emph{12},
  7127--7133\relax
\mciteBstWouldAddEndPuncttrue
\mciteSetBstMidEndSepPunct{\mcitedefaultmidpunct}
{\mcitedefaultendpunct}{\mcitedefaultseppunct}\relax
\EndOfBibitem
\bibitem[Noh \latin{et~al.}(2018)Noh, Choi, Kim, Im, Kim, Seo, and
  Lee]{Noh2018}
Noh,~G.; Choi,~D.; Kim,~J.-H.; Im,~D.-G.; Kim,~Y.-H.; Seo,~H.; Lee,~J. {Stark
  Tuning of Single-Photon Emitters in Hexagonal Boron Nitride}. \emph{Nano
  Lett.} \textbf{2018}, \emph{18}, 4710--4715\relax
\mciteBstWouldAddEndPuncttrue
\mciteSetBstMidEndSepPunct{\mcitedefaultmidpunct}
{\mcitedefaultendpunct}{\mcitedefaultseppunct}\relax
\EndOfBibitem
\bibitem[Nikolay \latin{et~al.}(2019)Nikolay, Mendelson, Sadzak, B{\"o}hm,
  Tran, Sontheimer, Aharonovich, and Benson]{Nikolay2019a}
Nikolay,~N.; Mendelson,~N.; Sadzak,~N.; B{\"o}hm,~F.; Tran,~T.~T.;
  Sontheimer,~B.; Aharonovich,~I.; Benson,~O. {Very Large and Reversible
  Stark-Shift Tuning of Single Emitters in Layered Hexagonal Boron Nitride}.
  \emph{Phys. Rev. Appl.} \textbf{2019}, \emph{11}, 041001\relax
\mciteBstWouldAddEndPuncttrue
\mciteSetBstMidEndSepPunct{\mcitedefaultmidpunct}
{\mcitedefaultendpunct}{\mcitedefaultseppunct}\relax
\EndOfBibitem
\bibitem[Xia \latin{et~al.}(2019)Xia, Li, Kim, Bao, Gong, Yang, Wang, and
  Zhang]{Xia2019}
Xia,~Y.; Li,~Q.; Kim,~J.; Bao,~W.; Gong,~C.; Yang,~S.; Wang,~Y.; Zhang,~X.
  {Room-Temperature Giant Stark Effect of Single Photon Emitter in Van Der
  Waals Material}. \emph{Nano Lett.} \textbf{2019}, \emph{19}, 7100--7105\relax
\mciteBstWouldAddEndPuncttrue
\mciteSetBstMidEndSepPunct{\mcitedefaultmidpunct}
{\mcitedefaultendpunct}{\mcitedefaultseppunct}\relax
\EndOfBibitem
\bibitem[Exarhos \latin{et~al.}(2019)Exarhos, Hopper, Patel, Doherty, and
  Bassett]{Exarhos2019}
Exarhos,~A.~L.; Hopper,~D.~A.; Patel,~R.~N.; Doherty,~M.~W.; Bassett,~L.~C.
  {Magnetic-Field-Dependent Quantum Emission in Hexagonal Boron Nitride at Room
  Temperature}. \emph{Nat. Commun.} \textbf{2019}, \emph{10}, 222\relax
\mciteBstWouldAddEndPuncttrue
\mciteSetBstMidEndSepPunct{\mcitedefaultmidpunct}
{\mcitedefaultendpunct}{\mcitedefaultseppunct}\relax
\EndOfBibitem
\bibitem[Chejanovsky \latin{et~al.}(2021)Chejanovsky, Mukherjee, Geng, Chen,
  Kim, Denisenko, Finkler, Taniguchi, Watanabe, Dasari, Auburger, Gali, Smet,
  and Wrachtrup]{Chejanovsky2021}
Chejanovsky,~N.; Mukherjee,~A.; Geng,~J.; Chen,~Y.-C.; Kim,~Y.; Denisenko,~A.;
  Finkler,~A.; Taniguchi,~T.; Watanabe,~K.; Dasari,~D. B.~R.; Auburger,~P.;
  Gali,~A.; Smet,~J.~H.; Wrachtrup,~J. {Single-Spin Resonance in a Van Der
  Waals Embedded Paramagnetic Defect}. \emph{Nat. Mater.} \textbf{2021},
  \emph{20}, 1079--1084\relax
\mciteBstWouldAddEndPuncttrue
\mciteSetBstMidEndSepPunct{\mcitedefaultmidpunct}
{\mcitedefaultendpunct}{\mcitedefaultseppunct}\relax
\EndOfBibitem
\bibitem[Stern \latin{et~al.}(2022)Stern, Gu, Jarman, Eizagirre~Barker,
  Mendelson, Chugh, Schott, Tan, Sirringhaus, Aharonovich, and
  Atat{\"u}re]{Stern2022}
Stern,~H.~L.; Gu,~Q.; Jarman,~J.; Eizagirre~Barker,~S.; Mendelson,~N.;
  Chugh,~D.; Schott,~S.; Tan,~H.~H.; Sirringhaus,~H.; Aharonovich,~I.;
  Atat{\"u}re,~M. {Room-Temperature Optically Detected Magnetic Resonance of
  Single Defects in Hexagonal Boron Nitride}. \emph{Nat. Commun.}
  \textbf{2022}, \emph{13}, 618\relax
\mciteBstWouldAddEndPuncttrue
\mciteSetBstMidEndSepPunct{\mcitedefaultmidpunct}
{\mcitedefaultendpunct}{\mcitedefaultseppunct}\relax
\EndOfBibitem
\bibitem[Abdi \latin{et~al.}(2018)Abdi, Chou, Gali, and Plenio]{Abdi2018}
Abdi,~M.; Chou,~J.-P.; Gali,~A.; Plenio,~M.~B. {Color Centers in Hexagonal
  Boron Nitride Monolayers: a Group Theory and Ab Initio Analysis}. \emph{ACS
  Photonics} \textbf{2018}, \emph{5}, 1967--1976\relax
\mciteBstWouldAddEndPuncttrue
\mciteSetBstMidEndSepPunct{\mcitedefaultmidpunct}
{\mcitedefaultendpunct}{\mcitedefaultseppunct}\relax
\EndOfBibitem
\bibitem[Wang \latin{et~al.}(2018)Wang, Zhang, Zhao, Luo, Wong, Wang, Wan,
  Venkatesan, Pennycook, Loh, Eda, and Wee]{Wang2018}
Wang,~Q.; Zhang,~Q.; Zhao,~X.; Luo,~X.; Wong,~C. P.~Y.; Wang,~J.; Wan,~D.;
  Venkatesan,~T.; Pennycook,~S.~J.; Loh,~K.~P.; Eda,~G.; Wee,~A. T.~S.
  {Photoluminescence Upconversion by Defects in Hexagonal Boron Nitride}.
  \emph{Nano Lett.} \textbf{2018}, \emph{18}, 6898--6905\relax
\mciteBstWouldAddEndPuncttrue
\mciteSetBstMidEndSepPunct{\mcitedefaultmidpunct}
{\mcitedefaultendpunct}{\mcitedefaultseppunct}\relax
\EndOfBibitem
\bibitem[Korona and Chojecki(2019)Korona, and Chojecki]{Korona2019}
Korona,~T.; Chojecki,~M. {Exploring Point Defects in Hexagonal Boron-Nitrogen
  Monolayers}. \emph{Int. J. Quantum Chem.} \textbf{2019}, \emph{119},
  e25925\relax
\mciteBstWouldAddEndPuncttrue
\mciteSetBstMidEndSepPunct{\mcitedefaultmidpunct}
{\mcitedefaultendpunct}{\mcitedefaultseppunct}\relax
\EndOfBibitem
\bibitem[Jara \latin{et~al.}(2021)Jara, Rauch, Botti, Marques, Norambuena,
  Coto, Castellanos-\'{A}guila, Maze, and Munoz]{Jara2021}
Jara,~C.; Rauch,~T.; Botti,~S.; Marques,~M.~A.; Norambuena,~A.; Coto,~R.;
  Castellanos-\'{A}guila,~J.; Maze,~J.; Munoz,~F. {First-Principles
  Identification of Single Photon Emitters Based on Carbon Clusters in
  Hexagonal Boron Nitride}. \emph{J. Phys. Chem. A} \textbf{2021}, \emph{125},
  1325--1335\relax
\mciteBstWouldAddEndPuncttrue
\mciteSetBstMidEndSepPunct{\mcitedefaultmidpunct}
{\mcitedefaultendpunct}{\mcitedefaultseppunct}\relax
\EndOfBibitem
\bibitem[Auburger and Gali(2021)Auburger, and Gali]{Auburger2021}
Auburger,~P.; Gali,~A. {Towards Ab Initio Identification of Paramagnetic
  Substitutional Carbon Defects in Hexagonal Boron Nitride Acting as Quantum
  Bits}. \emph{Phys. Rev. B} \textbf{2021}, \emph{104}, 075410\relax
\mciteBstWouldAddEndPuncttrue
\mciteSetBstMidEndSepPunct{\mcitedefaultmidpunct}
{\mcitedefaultendpunct}{\mcitedefaultseppunct}\relax
\EndOfBibitem
\bibitem[Tawfik \latin{et~al.}(2017)Tawfik, Ali, Fronzi, Kianinia, Tran,
  Stampfl, Aharonovich, Toth, and Ford]{Tawfik2017}
Tawfik,~S.~A.; Ali,~S.; Fronzi,~M.; Kianinia,~M.; Tran,~T.~T.; Stampfl,~C.;
  Aharonovich,~I.; Toth,~M.; Ford,~M.~J. {First-Principles Investigation of
  Quantum Emission from hBN Defects}. \emph{Nanoscale} \textbf{2017}, \emph{9},
  13575--13582\relax
\mciteBstWouldAddEndPuncttrue
\mciteSetBstMidEndSepPunct{\mcitedefaultmidpunct}
{\mcitedefaultendpunct}{\mcitedefaultseppunct}\relax
\EndOfBibitem
\bibitem[Sajid \latin{et~al.}(2018)Sajid, Reimers, and Ford]{Sajid2018}
Sajid,~A.; Reimers,~J.~R.; Ford,~M.~J. {Defect States in Hexagonal Boron
  Nitride: Assignments of Observed Properties and Prediction of Properties
  Relevant to Quantum Computation}. \emph{Phys. Rev. B} \textbf{2018},
  \emph{97}, 064101\relax
\mciteBstWouldAddEndPuncttrue
\mciteSetBstMidEndSepPunct{\mcitedefaultmidpunct}
{\mcitedefaultendpunct}{\mcitedefaultseppunct}\relax
\EndOfBibitem
\bibitem[Sajid and Thygesen(2020)Sajid, and Thygesen]{Sajid2020b}
Sajid,~A.; Thygesen,~K.~S. {$\mathrm{V}_\mathrm{N} \mathrm{C}_\mathrm{B}$
  Defect as Source of Single Photon Emission from Hexagonal Boron Nitride}.
  \emph{2D Mater.} \textbf{2020}, \emph{7}, 031007\relax
\mciteBstWouldAddEndPuncttrue
\mciteSetBstMidEndSepPunct{\mcitedefaultmidpunct}
{\mcitedefaultendpunct}{\mcitedefaultseppunct}\relax
\EndOfBibitem
\bibitem[Weston \latin{et~al.}(2018)Weston, Wickramaratne, Mackoit, Alkauskas,
  and Van~de Walle]{Weston2018}
Weston,~L.; Wickramaratne,~D.; Mackoit,~M.; Alkauskas,~A.; Van~de Walle,~C.~G.
  {Native Point Defects and Impurities in Hexagonal Boron Nitride}. \emph{Phys.
  Rev. B} \textbf{2018}, \emph{97}, 214104\relax
\mciteBstWouldAddEndPuncttrue
\mciteSetBstMidEndSepPunct{\mcitedefaultmidpunct}
{\mcitedefaultendpunct}{\mcitedefaultseppunct}\relax
\EndOfBibitem
\bibitem[Turiansky \latin{et~al.}(2019)Turiansky, Alkauskas, Bassett, and
  Van~de Walle]{Turiansky2019}
Turiansky,~M.~E.; Alkauskas,~A.; Bassett,~L.~C.; Van~de Walle,~C.~G. {Dangling
  Bonds in Hexagonal Boron Nitride as Single-Photon Emitters}. \emph{Phys. Rev.
  Lett.} \textbf{2019}, \emph{123}, 127401\relax
\mciteBstWouldAddEndPuncttrue
\mciteSetBstMidEndSepPunct{\mcitedefaultmidpunct}
{\mcitedefaultendpunct}{\mcitedefaultseppunct}\relax
\EndOfBibitem
\bibitem[Lyu \latin{et~al.}(2020)Lyu, Zhu, Gu, Qiao, Watanabe, Taniguchi, and
  Ye]{Lyu2020}
Lyu,~C.; Zhu,~Y.; Gu,~P.; Qiao,~J.; Watanabe,~K.; Taniguchi,~T.; Ye,~Y.
  {Single-Photon Emission from Two-Dimensional Hexagonal Boron Nitride Annealed
  in a Carbon-Rich Environment}. \emph{Appl. Phys. Lett.} \textbf{2020},
  \emph{117}, 244002\relax
\mciteBstWouldAddEndPuncttrue
\mciteSetBstMidEndSepPunct{\mcitedefaultmidpunct}
{\mcitedefaultendpunct}{\mcitedefaultseppunct}\relax
\EndOfBibitem
\bibitem[Hayee \latin{et~al.}(2020)Hayee, Yu, Zhang, Ciccarino, Nguyen,
  Marshall, Aharonovich, Vu{\v{c}}kovi{\'c}, Narang, Heinz, and
  Dionne]{Hayee2020}
Hayee,~F.; Yu,~L.; Zhang,~J.~L.; Ciccarino,~C.~J.; Nguyen,~M.; Marshall,~A.~F.;
  Aharonovich,~I.; Vu{\v{c}}kovi{\'c},~J.; Narang,~P.; Heinz,~T.~F.;
  Dionne,~J.~A. {Revealing Multiple Classes of Stable Quantum Emitters in
  Hexagonal Boron Nitride with Correlated Optical and Electron Microscopy}.
  \emph{Nat. Mater.} \textbf{2020}, \emph{19}, 534--539\relax
\mciteBstWouldAddEndPuncttrue
\mciteSetBstMidEndSepPunct{\mcitedefaultmidpunct}
{\mcitedefaultendpunct}{\mcitedefaultseppunct}\relax
\EndOfBibitem
\bibitem[Norinaga \latin{et~al.}(2009)Norinaga, Deutschmann, Saegusa, and
  Hayashi]{Norinaga2009}
Norinaga,~K.; Deutschmann,~O.; Saegusa,~N.; Hayashi,~J.-I. {Analysis of
  Pyrolysis Products from Light Hydrocarbons and Kinetic Modeling for Growth of
  Polycyclic Aromatic Hydrocarbons with Detailed Chemistry}. \emph{J. Anal.
  Appl. Pyrolysis} \textbf{2009}, \emph{86}, 148--160\relax
\mciteBstWouldAddEndPuncttrue
\mciteSetBstMidEndSepPunct{\mcitedefaultmidpunct}
{\mcitedefaultendpunct}{\mcitedefaultseppunct}\relax
\EndOfBibitem
\bibitem[Viteri \latin{et~al.}(2019)Viteri, L{\'o}pez, Millera, Bilbao, and
  Alzueta]{Viteri2019}
Viteri,~F.; L{\'o}pez,~A.; Millera,~{\'A}.; Bilbao,~R.; Alzueta,~M.~U.
  {Influence of Temperature and Gas Residence Time on the Formation of
  Polycyclic Aromatic Hydrocarbons (PAH) during the Pyrolysis of Ethanol}.
  \emph{Fuel} \textbf{2019}, \emph{236}, 820--828\relax
\mciteBstWouldAddEndPuncttrue
\mciteSetBstMidEndSepPunct{\mcitedefaultmidpunct}
{\mcitedefaultendpunct}{\mcitedefaultseppunct}\relax
\EndOfBibitem
\bibitem[Zhou \latin{et~al.}(2015)Zhou, Wu, Onwudili, Meng, Zhang, and
  Williams]{Zhou2015}
Zhou,~H.; Wu,~C.; Onwudili,~J.~A.; Meng,~A.; Zhang,~Y.; Williams,~P.~T.
  {Polycyclic Aromatic Hydrocarbons (PAH) Formation from the Pyrolysis of
  Different Municipal Solid Waste Fractions}. \emph{Waste Management}
  \textbf{2015}, \emph{36}, 136--146\relax
\mciteBstWouldAddEndPuncttrue
\mciteSetBstMidEndSepPunct{\mcitedefaultmidpunct}
{\mcitedefaultendpunct}{\mcitedefaultseppunct}\relax
\EndOfBibitem
\bibitem[Rieger and M{\"u}llen(2010)Rieger, and M{\"u}llen]{Rieger2010}
Rieger,~R.; M{\"u}llen,~K. {Forever Young: Polycyclic Aromatic Hydrocarbons as
  Model Cases for Structural and Optical Studies}. \emph{J. Phys. Org. Chem.}
  \textbf{2010}, \emph{23}, 315--325\relax
\mciteBstWouldAddEndPuncttrue
\mciteSetBstMidEndSepPunct{\mcitedefaultmidpunct}
{\mcitedefaultendpunct}{\mcitedefaultseppunct}\relax
\EndOfBibitem
\bibitem[Toninelli \latin{et~al.}(2021)Toninelli, Gerhardt, Clark,
  Reserbat-Plantey, G{\"o}tzinger, Ristanovi{\'c}, Colautti, Lombardi, Major,
  Deperasi{\'n}ska, Pernice, Koppens, Kozankiewicz, Gourdon, Sandoghdar, and
  Orrit]{Toninelli2021}
Toninelli,~C.; Gerhardt,~I.; Clark,~A.~S.; Reserbat-Plantey,~A.;
  G{\"o}tzinger,~S.; Ristanovi{\'c},~Z.; Colautti,~M.; Lombardi,~P.;
  Major,~K.~D.; Deperasi{\'n}ska,~I.; Pernice,~W.; Koppens,~F.;
  Kozankiewicz,~B.; Gourdon,~A.; Sandoghdar,~V.; Orrit,~M. {Single Organic
  Molecules for Photonic Quantum Technologies}. \emph{Nat. Mater.}
  \textbf{2021}, \emph{20}, 1615--–1628\relax
\mciteBstWouldAddEndPuncttrue
\mciteSetBstMidEndSepPunct{\mcitedefaultmidpunct}
{\mcitedefaultendpunct}{\mcitedefaultseppunct}\relax
\EndOfBibitem
\bibitem[Watanabe \latin{et~al.}(2004)Watanabe, Taniguchi, and
  Kanda]{Watanabe2004}
Watanabe,~K.; Taniguchi,~T.; Kanda,~H. {Direct-Bandgap Properties and Evidence
  for Ultraviolet Lasing of Hexagonal Boron Nitride Single Crystal}. \emph{Nat.
  Mater.} \textbf{2004}, \emph{3}, 404--409\relax
\mciteBstWouldAddEndPuncttrue
\mciteSetBstMidEndSepPunct{\mcitedefaultmidpunct}
{\mcitedefaultendpunct}{\mcitedefaultseppunct}\relax
\EndOfBibitem
\bibitem[Gasparutti \latin{et~al.}(2020)Gasparutti, Song, Neumann, Wei,
  Watanabe, Taniguchi, and Lee]{Gasparutti2020}
Gasparutti,~I.; Song,~S.~H.; Neumann,~M.; Wei,~X.; Watanabe,~K.; Taniguchi,~T.;
  Lee,~Y.~H. {How Clean Is Clean? Recipes for Van Der Waals Heterostructure
  Cleanliness Assessment}. \emph{ACS Appl. Mater. Interfaces} \textbf{2020},
  \emph{12}, 7701--7709\relax
\mciteBstWouldAddEndPuncttrue
\mciteSetBstMidEndSepPunct{\mcitedefaultmidpunct}
{\mcitedefaultendpunct}{\mcitedefaultseppunct}\relax
\EndOfBibitem
\bibitem[Kulzer \latin{et~al.}(1999)Kulzer, Koberling, Christ, Mews, and
  Basch{\'e}]{Kulzer1999a}
Kulzer,~F.; Koberling,~F.; Christ,~T.; Mews,~A.; Basch{\'e},~T. {Terrylene in
  p-Terphenyl: Single-Molecule Experiments at Room Temperature}. \emph{Chem.
  Phys.} \textbf{1999}, \emph{247}, 23--34\relax
\mciteBstWouldAddEndPuncttrue
\mciteSetBstMidEndSepPunct{\mcitedefaultmidpunct}
{\mcitedefaultendpunct}{\mcitedefaultseppunct}\relax
\EndOfBibitem
\bibitem[Chen \latin{et~al.}(2019)Chen, Thoms, St{\"o}ttinger, Schollmeyer,
  M{\"u}llen, Narita, and Basch{\'e}]{Chen2019}
Chen,~Q.; Thoms,~S.; St{\"o}ttinger,~S.; Schollmeyer,~D.; M{\"u}llen,~K.;
  Narita,~A.; Basch{\'e},~T. {Dibenzo[{\it hi,st}]ovalene as Highly Luminescent
  Nanographene: Efficient Synthesis via Photochemical Cyclodehydroiodination,
  Optoelectronic Properties, and Single-Molecule Spectroscopy}. \emph{J. Am.
  Chem. Soc.} \textbf{2019}, \emph{141}, 16439--16449\relax
\mciteBstWouldAddEndPuncttrue
\mciteSetBstMidEndSepPunct{\mcitedefaultmidpunct}
{\mcitedefaultendpunct}{\mcitedefaultseppunct}\relax
\EndOfBibitem
\bibitem[Garcia \latin{et~al.}(2012)Garcia, Neumann, Amet, Williams, Watanabe,
  Taniguchi, and Goldhaber-Gordon]{Garcia2012}
Garcia,~A. G.~F.; Neumann,~M.; Amet,~F.; Williams,~J.~R.; Watanabe,~K.;
  Taniguchi,~T.; Goldhaber-Gordon,~D. {Effective Cleaning of Hexagonal Boron
  Nitride for Graphene Devices}. \emph{Nano Lett.} \textbf{2012}, \emph{12},
  4449--4454\relax
\mciteBstWouldAddEndPuncttrue
\mciteSetBstMidEndSepPunct{\mcitedefaultmidpunct}
{\mcitedefaultendpunct}{\mcitedefaultseppunct}\relax
\EndOfBibitem
\bibitem[Buechner(1975)]{Buechner1975}
Buechner,~U. {The Dielectric Function of Mica and Quartz Determined by Electron
  Energy Losses}. \emph{J. Phys. C: Solid State Phys.} \textbf{1975}, \emph{8},
  2781--2787\relax
\mciteBstWouldAddEndPuncttrue
\mciteSetBstMidEndSepPunct{\mcitedefaultmidpunct}
{\mcitedefaultendpunct}{\mcitedefaultseppunct}\relax
\EndOfBibitem
\bibitem[G{\"u}ttler \latin{et~al.}(1996)G{\"u}ttler, Croci, Renn, and
  Wild]{Guttler1996}
G{\"u}ttler,~F.; Croci,~M.; Renn,~A.; Wild,~U.~P. {Single Molecule Polarization
  Spectroscopy: Pentacene in p-Terphenyl}. \emph{Chem. Phys.} \textbf{1996},
  \emph{211}, 421--430\relax
\mciteBstWouldAddEndPuncttrue
\mciteSetBstMidEndSepPunct{\mcitedefaultmidpunct}
{\mcitedefaultendpunct}{\mcitedefaultseppunct}\relax
\EndOfBibitem
\bibitem[Pfab \latin{et~al.}(2004)Pfab, Zimmermann, Hettich, Gerhardt, Renn,
  and Sandoghdar]{Pfab2004}
Pfab,~R.~J.; Zimmermann,~J.; Hettich,~C.; Gerhardt,~I.; Renn,~A.;
  Sandoghdar,~V. {Aligned Terrylene Molecules in a Spin-Coated Ultrathin
  Crystalline Film of p-Terphenyl}. \emph{Chem. Phys. Lett.} \textbf{2004},
  \emph{387}, 490--495\relax
\mciteBstWouldAddEndPuncttrue
\mciteSetBstMidEndSepPunct{\mcitedefaultmidpunct}
{\mcitedefaultendpunct}{\mcitedefaultseppunct}\relax
\EndOfBibitem
\bibitem[Nicolet \latin{et~al.}(2007)Nicolet, Bordat, Hofmann, Kol'chenko,
  Kozankiewicz, Brown, and Orrit]{Nicolet2007}
Nicolet,~A.~A.; Bordat,~P.; Hofmann,~C.; Kol'chenko,~M.~A.; Kozankiewicz,~B.;
  Brown,~R.; Orrit,~M. {Single Dibenzoterrylene Molecules in an Anthracene
  Crystal: Main Insertion Sites}. \emph{ChemPhysChem} \textbf{2007}, \emph{8},
  1929--1936\relax
\mciteBstWouldAddEndPuncttrue
\mciteSetBstMidEndSepPunct{\mcitedefaultmidpunct}
{\mcitedefaultendpunct}{\mcitedefaultseppunct}\relax
\EndOfBibitem
\bibitem[Y{\"u}ce and Kiraz(2012)Y{\"u}ce, and Kiraz]{Yuce2012}
Y{\"u}ce,~M.~Y.; Kiraz,~A. {Single-Molecule Fluorescence of Terrylene Embedded
  in Anthracene Matrix: a Room Temperature Study}. \emph{Chem. Phys. Lett.}
  \textbf{2012}, \emph{547}, 47--51\relax
\mciteBstWouldAddEndPuncttrue
\mciteSetBstMidEndSepPunct{\mcitedefaultmidpunct}
{\mcitedefaultendpunct}{\mcitedefaultseppunct}\relax
\EndOfBibitem
\bibitem[Croci \latin{et~al.}(1993)Croci, M{\"u}schenborn, G{\"u}ttler, Renn,
  and Wild]{Croci1993}
Croci,~M.; M{\"u}schenborn,~H.-J.; G{\"u}ttler,~F.; Renn,~A.; Wild,~U.~P.
  {Single Molecule Spectroscopy: Pressure Effect on Pentacene in p-Terphenyl}.
  \emph{Chem. Phys. Lett.} \textbf{1993}, \emph{212}, 71--77\relax
\mciteBstWouldAddEndPuncttrue
\mciteSetBstMidEndSepPunct{\mcitedefaultmidpunct}
{\mcitedefaultendpunct}{\mcitedefaultseppunct}\relax
\EndOfBibitem
\bibitem[M{\"u}ller \latin{et~al.}(1995)M{\"u}ller, Richter, and
  Kador]{Muller1995}
M{\"u}ller,~A.; Richter,~W.; Kador,~L. {Pressure Effects on Single Molecules of
  Terrylene in p-Terphenyl}. \emph{Chem. Phys. Lett.} \textbf{1995},
  \emph{241}, 547--554\relax
\mciteBstWouldAddEndPuncttrue
\mciteSetBstMidEndSepPunct{\mcitedefaultmidpunct}
{\mcitedefaultendpunct}{\mcitedefaultseppunct}\relax
\EndOfBibitem
\bibitem[Iwamoto \latin{et~al.}(1998)Iwamoto, Kurita, and Kushida]{Iwamoto1998}
Iwamoto,~T.; Kurita,~A.; Kushida,~T. {Pressure Effects on Single-Molecule
  Spectra of Terrylene in Hexadecane}. \emph{Chem. Phys. Lett.} \textbf{1998},
  \emph{284}, 147--152\relax
\mciteBstWouldAddEndPuncttrue
\mciteSetBstMidEndSepPunct{\mcitedefaultmidpunct}
{\mcitedefaultendpunct}{\mcitedefaultseppunct}\relax
\EndOfBibitem
\bibitem[Wild \latin{et~al.}(1992)Wild, G{\"u}ttler, Pirotta, and
  Renn]{Wild1992}
Wild,~U.~P.; G{\"u}ttler,~F.; Pirotta,~M.; Renn,~A. {Single Molecule
  Spectroscopy: Stark Effect of Pentacene in p-Terphenyl}. \emph{Chem. Phys.
  Lett.} \textbf{1992}, \emph{193}, 451--455\relax
\mciteBstWouldAddEndPuncttrue
\mciteSetBstMidEndSepPunct{\mcitedefaultmidpunct}
{\mcitedefaultendpunct}{\mcitedefaultseppunct}\relax
\EndOfBibitem
\bibitem[Orrit \latin{et~al.}(1992)Orrit, Bernard, Zumbusch, and
  Personov]{Orrit1992}
Orrit,~M.; Bernard,~J.; Zumbusch,~A.; Personov,~R.~I. {Stark Effect on Single
  Molecules in a Polymer Matrix}. \emph{Chem. Phys. Lett.} \textbf{1992},
  \emph{196}, 595--600\relax
\mciteBstWouldAddEndPuncttrue
\mciteSetBstMidEndSepPunct{\mcitedefaultmidpunct}
{\mcitedefaultendpunct}{\mcitedefaultseppunct}\relax
\EndOfBibitem
\bibitem[Kulzer \latin{et~al.}(1999)Kulzer, Matzke, Br{\"a}uchle, and
  Basch{\'e}]{Kulzer1999b}
Kulzer,~F.; Matzke,~R.; Br{\"a}uchle,~C.; Basch{\'e},~T. {Nonphotochemical Hole
  Burning Investigated at the Single-Molecule Level: Stark Effect Measurements
  on the Original and Photoproduct State}. \emph{J. Phys. Chem. A}
  \textbf{1999}, \emph{103}, 2408--2411\relax
\mciteBstWouldAddEndPuncttrue
\mciteSetBstMidEndSepPunct{\mcitedefaultmidpunct}
{\mcitedefaultendpunct}{\mcitedefaultseppunct}\relax
\EndOfBibitem
\bibitem[Bauer and Kador(2003)Bauer, and Kador]{Bauer2003}
Bauer,~M.; Kador,~L. {Electric-Field Effects of Two-Level Systems Observed with
  Single-Molecule Spectroscopy}. \emph{J. Chem. Phys.} \textbf{2003},
  \emph{118}, 9069--9072\relax
\mciteBstWouldAddEndPuncttrue
\mciteSetBstMidEndSepPunct{\mcitedefaultmidpunct}
{\mcitedefaultendpunct}{\mcitedefaultseppunct}\relax
\EndOfBibitem
\bibitem[Moradi \latin{et~al.}(2019)Moradi, Ristanovi{\'c}, Orrit,
  Deperasi{\'n}ska, and Kozankiewicz]{Moradi2019}
Moradi,~A.; Ristanovi{\'c},~Z.; Orrit,~M.; Deperasi{\'n}ska,~I.;
  Kozankiewicz,~B. {Matrix-Induced Linear Stark Effect of Single
  Dibenzoterrylene Molecules in 2,3-Dibromonaphthalene Crystal}.
  \emph{ChemPhysChem} \textbf{2019}, \emph{20}, 55--61\relax
\mciteBstWouldAddEndPuncttrue
\mciteSetBstMidEndSepPunct{\mcitedefaultmidpunct}
{\mcitedefaultendpunct}{\mcitedefaultseppunct}\relax
\EndOfBibitem
\bibitem[Gerson and Huber(2003)Gerson, and Huber]{Gerson2003}
Gerson,~F.; Huber,~W. \emph{{Electron Spin Resonance Spectroscopy of Organic
  Radicals}}; Wiley-VCH: Weinheim, 2003; pp 10--36, 97--165, 210--289\relax
\mciteBstWouldAddEndPuncttrue
\mciteSetBstMidEndSepPunct{\mcitedefaultmidpunct}
{\mcitedefaultendpunct}{\mcitedefaultseppunct}\relax
\EndOfBibitem
\bibitem[Morita \latin{et~al.}(2011)Morita, Suzuki, Sato, and
  Takui]{Morita2011}
Morita,~Y.; Suzuki,~S.; Sato,~K.; Takui,~T. {Synthetic Organic Spin Chemistry
  for Structurally Well-Defined Open-Shell Graphene Fragments}. \emph{Nat.
  Chem.} \textbf{2011}, \emph{3}, 197--204\relax
\mciteBstWouldAddEndPuncttrue
\mciteSetBstMidEndSepPunct{\mcitedefaultmidpunct}
{\mcitedefaultendpunct}{\mcitedefaultseppunct}\relax
\EndOfBibitem
\bibitem[Sun and Wu(2012)Sun, and Wu]{Sun2012}
Sun,~Z.; Wu,~J. {Open-Shell Polycyclic Aromatic Hydrocarbons}. \emph{J. Mater.
  Chem.} \textbf{2012}, \emph{22}, 4151--4160\relax
\mciteBstWouldAddEndPuncttrue
\mciteSetBstMidEndSepPunct{\mcitedefaultmidpunct}
{\mcitedefaultendpunct}{\mcitedefaultseppunct}\relax
\EndOfBibitem
\bibitem[Ahmed and Mandal(2022)Ahmed, and Mandal]{Ahmed2022}
Ahmed,~J.; Mandal,~S.~K. {Phenalenyl Radical: Smallest Polycyclic Odd Alternant
  Hydrocarbon Present in the Graphene Sheet}. \emph{Chem. Rev.} \textbf{2022},
  \emph{122}, 11369--11431\relax
\mciteBstWouldAddEndPuncttrue
\mciteSetBstMidEndSepPunct{\mcitedefaultmidpunct}
{\mcitedefaultendpunct}{\mcitedefaultseppunct}\relax
\EndOfBibitem
\bibitem[Cofino \latin{et~al.}(1984)Cofino, van Dam, Kamminga, Hoornweg,
  Gooijer, MacLean, and Velthorst]{Cofino1984}
Cofino,~W.~P.; van Dam,~S.~M.; Kamminga,~D.~A.; Hoornweg,~G.~P.; Gooijer,~C.;
  MacLean,~C.; Velthorst,~N.~H. {Jahn-Teller Effect in Highly Resolved Optical
  Spectra of the Phenalenyl Radical}. \emph{Mol. Phys.} \textbf{1984},
  \emph{51}, 537--550\relax
\mciteBstWouldAddEndPuncttrue
\mciteSetBstMidEndSepPunct{\mcitedefaultmidpunct}
{\mcitedefaultendpunct}{\mcitedefaultseppunct}\relax
\EndOfBibitem
\bibitem[O'Connor \latin{et~al.}(2011)O'Connor, Troy, Roberts, Chalyavi,
  F{\"u}ckel, Crossley, Nauta, Stanton, and Schmidt]{OConnor2011}
O'Connor,~G.~D.; Troy,~T.~P.; Roberts,~D.~A.; Chalyavi,~N.; F{\"u}ckel,~B.;
  Crossley,~M.~J.; Nauta,~K.; Stanton,~J.~F.; Schmidt,~T.~W. {Spectroscopy of
  the Free Phenalenyl Radical}. \emph{J. Am. Chem. Soc.} \textbf{2011},
  \emph{133}, 14554--14557\relax
\mciteBstWouldAddEndPuncttrue
\mciteSetBstMidEndSepPunct{\mcitedefaultmidpunct}
{\mcitedefaultendpunct}{\mcitedefaultseppunct}\relax
\EndOfBibitem
\bibitem[Pavli{\v{c}}ek \latin{et~al.}(2017)Pavli{\v{c}}ek, Mistry, Majzik,
  Moll, Meyer, Fox, and Gross]{Pavlicek2017}
Pavli{\v{c}}ek,~N.; Mistry,~A.; Majzik,~Z.; Moll,~N.; Meyer,~G.; Fox,~D.~J.;
  Gross,~L. {Synthesis and Characterization of Triangulene}. \emph{Nat.
  Nanotechnol.} \textbf{2017}, \emph{12}, 308--311\relax
\mciteBstWouldAddEndPuncttrue
\mciteSetBstMidEndSepPunct{\mcitedefaultmidpunct}
{\mcitedefaultendpunct}{\mcitedefaultseppunct}\relax
\EndOfBibitem
\bibitem[Mishra \latin{et~al.}(2019)Mishra, Beyer, Eimre, Liu, Berger,
  Gr{\"o}ning, Pignedoli, M{\"u}llen, Fasel, Feng, and Ruffieux]{Mishra2019}
Mishra,~S.; Beyer,~D.; Eimre,~K.; Liu,~J.; Berger,~R.; Gr{\"o}ning,~O.;
  Pignedoli,~C.~A.; M{\"u}llen,~K.; Fasel,~R.; Feng,~X.; Ruffieux,~P.
  {Synthesis and Characterization of $\pi$-Extended Triangulene}. \emph{J. Am.
  Chem. Soc.} \textbf{2019}, \emph{141}, 10621--10625\relax
\mciteBstWouldAddEndPuncttrue
\mciteSetBstMidEndSepPunct{\mcitedefaultmidpunct}
{\mcitedefaultendpunct}{\mcitedefaultseppunct}\relax
\EndOfBibitem
\bibitem[Khatri \latin{et~al.}(2019)Khatri, Luxmoore, and Ramsay]{Khatri2019}
Khatri,~P.; Luxmoore,~I.~J.; Ramsay,~A.~J. {Phonon Sidebands of Color Centers
  in Hexagonal Boron Nitride}. \emph{Phys. Rev. B} \textbf{2019}, \emph{100},
  125305\relax
\mciteBstWouldAddEndPuncttrue
\mciteSetBstMidEndSepPunct{\mcitedefaultmidpunct}
{\mcitedefaultendpunct}{\mcitedefaultseppunct}\relax
\EndOfBibitem
\bibitem[Wigger \latin{et~al.}(2019)Wigger, Schmidt, Del Pozo-Zamudio,
  Preu{\ss}, Tonndorf, Schneider, Steeger, Kern, Khodaei, Sperling,
  Michaelis~de Vasconcellos, Bratschitsch, and Kuhn]{Wigger2019}
Wigger,~D.; Schmidt,~R.; Del Pozo-Zamudio,~O.; Preu{\ss},~J.~A.; Tonndorf,~P.;
  Schneider,~R.; Steeger,~P.; Kern,~J.; Khodaei,~Y.; Sperling,~J.; Michaelis~de
  Vasconcellos,~S.; Bratschitsch,~R.; Kuhn,~T. {Phonon-Assisted Emission and
  Absorption of Individual Color Centers in Hexagonal Boron Nitride}. \emph{2D
  Mater.} \textbf{2019}, \emph{6}, 035006\relax
\mciteBstWouldAddEndPuncttrue
\mciteSetBstMidEndSepPunct{\mcitedefaultmidpunct}
{\mcitedefaultendpunct}{\mcitedefaultseppunct}\relax
\EndOfBibitem
\bibitem[Kumar \latin{et~al.}(2022)Kumar, Cholsuk, Zand, Mishuk, Matthes,
  Eilenberger, Suwanna, and Vogl]{Kumar2022}
Kumar,~A.; Cholsuk,~C.; Zand,~A.; Mishuk,~M.~N.; Matthes,~T.; Eilenberger,~F.;
  Suwanna,~S.; Vogl,~T. {Localized Creation of Yellow Single Photon Emitting
  Carbon Complexes in Hexagonal Boron Nitride}. \textbf{2022}, 2208.13488,
  arXiv, https://arxiv.org/abs/2208.13488\relax
\mciteBstWouldAddEndPuncttrue
\mciteSetBstMidEndSepPunct{\mcitedefaultmidpunct}
{\mcitedefaultendpunct}{\mcitedefaultseppunct}\relax
\EndOfBibitem
\bibitem[Kumar \latin{et~al.}(2023)Kumar, Samaner, Cholsuk, Matthes, Pa\c{c}al,
  Oyun, Zand, Chapman, Saerens, Grange, Suwanna, Ate\c{s}, and Vogl]{Kumar2023}
Kumar,~A.; Samaner,~C.; Cholsuk,~C.; Matthes,~T.; Pa\c{c}al,~S.; Oyun,~Y.;
  Zand,~A.; Chapman,~R.~J.; Saerens,~G.; Grange,~R.; Suwanna,~S.; Ate\c{s},~S.;
  Vogl,~T. {Polarization Dynamics of Solid-State Quantum Emitters}.
  \textbf{2023}, 2303.04732, arXiv, https://arxiv.org/abs/2303.04732\relax
\mciteBstWouldAddEndPuncttrue
\mciteSetBstMidEndSepPunct{\mcitedefaultmidpunct}
{\mcitedefaultendpunct}{\mcitedefaultseppunct}\relax
\EndOfBibitem
\bibitem[Taniguchi and Watanabe(2007)Taniguchi, and Watanabe]{Taniguchi2007}
Taniguchi,~T.; Watanabe,~K. {Synthesis of High-Purity Boron Nitride Single
  Crystals under High Pressure by Using Ba-BN Solvent}. \emph{J. Cryst. Growth}
  \textbf{2007}, \emph{303}, 525--529\relax
\mciteBstWouldAddEndPuncttrue
\mciteSetBstMidEndSepPunct{\mcitedefaultmidpunct}
{\mcitedefaultendpunct}{\mcitedefaultseppunct}\relax
\EndOfBibitem
\bibitem[Kuo \latin{et~al.}(2016)Kuo, Neumann, Balamurugan, Park, Kang, Shiu,
  Kang, Hong, Han, Noh, and Park]{Kuo2016}
Kuo,~C.-T.; Neumann,~M.; Balamurugan,~K.; Park,~H.~J.; Kang,~S.; Shiu,~H.~W.;
  Kang,~J.~H.; Hong,~B.~H.; Han,~M.; Noh,~T.~W.; Park,~J.-G. {Exfoliation and
  Raman Spectroscopic Fingerprint of Few-Layer NiPS$_3$ Van Der Waals
  Crystals}. \emph{Sci. Rep.} \textbf{2016}, \emph{6}, 1--10\relax
\mciteBstWouldAddEndPuncttrue
\mciteSetBstMidEndSepPunct{\mcitedefaultmidpunct}
{\mcitedefaultendpunct}{\mcitedefaultseppunct}\relax
\EndOfBibitem
\bibitem[Li \latin{et~al.}(2014)Li, Cervenka, Watanabe, Taniguchi, and
  Chen]{Li2014}
Li,~L.~H.; Cervenka,~J.; Watanabe,~K.; Taniguchi,~T.; Chen,~Y. {Strong
  Oxidation Resistance of Atomically Thin Boron Nitride Nanosheets}. \emph{ACS
  Nano} \textbf{2014}, \emph{8}, 1457--1462\relax
\mciteBstWouldAddEndPuncttrue
\mciteSetBstMidEndSepPunct{\mcitedefaultmidpunct}
{\mcitedefaultendpunct}{\mcitedefaultseppunct}\relax
\EndOfBibitem
\bibitem[Hollrichter(2010)]{Hollrichter2010}
Hollrichter,~O. {Raman Instrumentation for Confocal Raman Microscopy}. In
  \emph{{Confocal Raman Microscopy}}; Dieing,~T., Hollrichter,~O.,
  Toporski,~J., Eds.; Springer: Berlin, Heidelberg, 2010; pp 43--60\relax
\mciteBstWouldAddEndPuncttrue
\mciteSetBstMidEndSepPunct{\mcitedefaultmidpunct}
{\mcitedefaultendpunct}{\mcitedefaultseppunct}\relax
\EndOfBibitem
\bibitem[Schindelin \latin{et~al.}(2012)Schindelin, Arganda-Carreras, Frise,
  Kaynig, Longair, Pietzsch, Preibisch, Rueden, Saalfeld, Schmid, Tinevez,
  White, Hartenstein, Eliceiri, Tomancak, and Cardona]{Schindelin2012}
Schindelin,~J.; Arganda-Carreras,~I.; Frise,~E.; Kaynig,~V.; Longair,~M.;
  Pietzsch,~T.; Preibisch,~S.; Rueden,~C.; Saalfeld,~S.; Schmid,~B.;
  Tinevez,~J.-Y.; White,~D.~J.; Hartenstein,~V.; Eliceiri,~K.; Tomancak,~P.;
  Cardona,~A. {Fiji: an Open-Source Platform for Biological-Image Analysis}.
  \emph{Nat. Methods} \textbf{2012}, \emph{9}, 676--682\relax
\mciteBstWouldAddEndPuncttrue
\mciteSetBstMidEndSepPunct{\mcitedefaultmidpunct}
{\mcitedefaultendpunct}{\mcitedefaultseppunct}\relax
\EndOfBibitem
\bibitem[Ne{\v{c}}as and Klapetek(2012)Ne{\v{c}}as, and Klapetek]{Necas2012}
Ne{\v{c}}as,~D.; Klapetek,~P. {Gwyddion: an Open-Source Software for SPM Data
  Analysis}. \emph{Cent. Eur. J. Phys} \textbf{2012}, \emph{10}, 181--188\relax
\mciteBstWouldAddEndPuncttrue
\mciteSetBstMidEndSepPunct{\mcitedefaultmidpunct}
{\mcitedefaultendpunct}{\mcitedefaultseppunct}\relax
\EndOfBibitem
\end{mcitethebibliography}

\clearpage


\begin{figure}[ht]
\begin{center}
\includegraphics[width=11.5 cm]{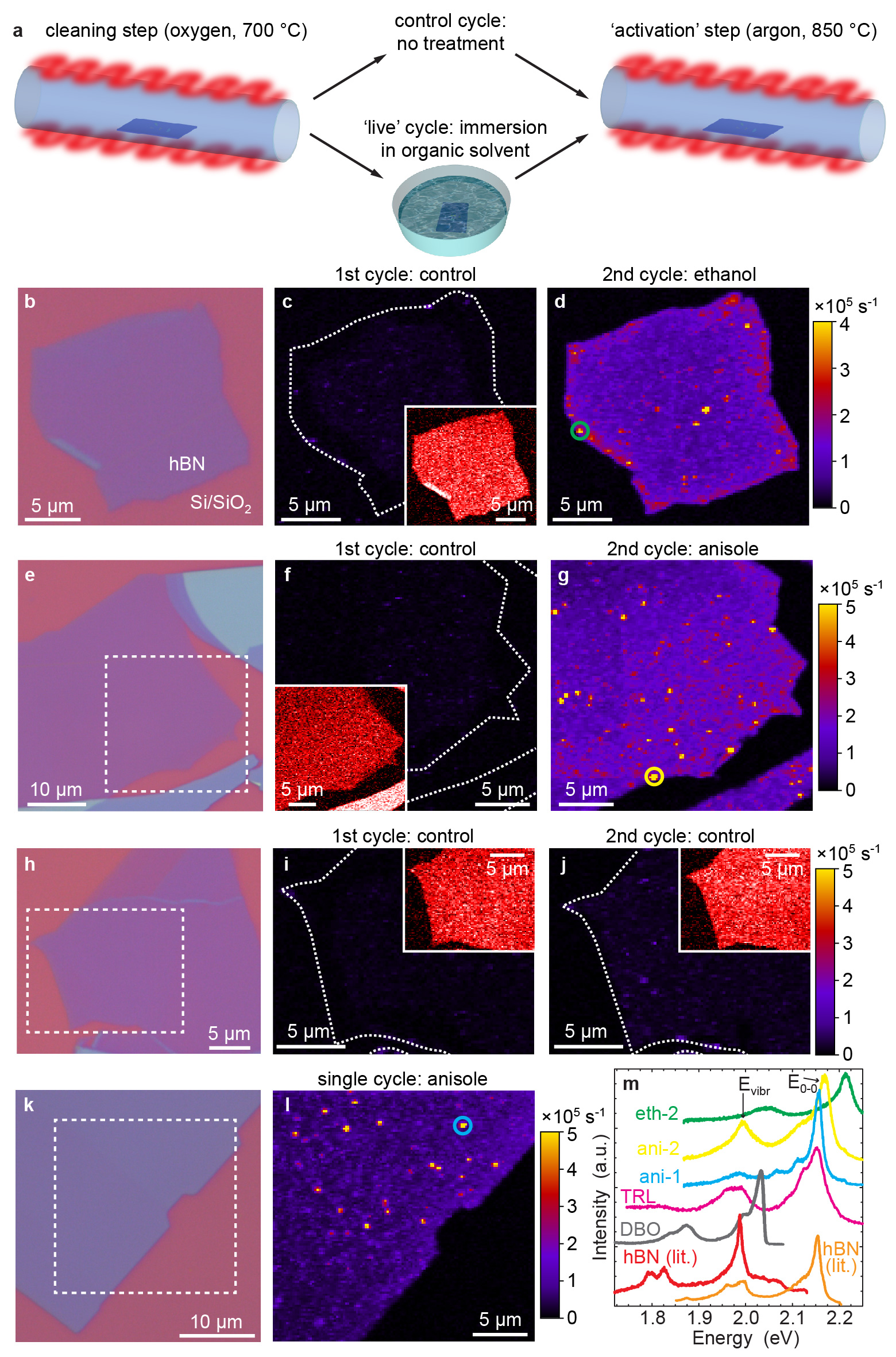}     
\end{center}
\end{figure}

Figure 1: \textbf{Emission centers at hBN sheets originate from organic impurities.} \textbf{a}, Schematic of the experimental
procedure. Control cycle: hBN samples are subjected to cleaning in oxygen atmosphere, followed by an activation heat treatment
in argon at 850 \degC. Live cycle: same, with immersion in organic solvent as an additional step inserted between oxidative
cleaning and activation, for the controlled introduction of organic residue. The key role of organic impurities is demonstrated
by consecutively performing a control cycle and live cycle on the same samples. \textbf{b}, Photograph of exfoliated hBN sheet
(thickness 12 nm). \textbf{c},\textbf{d}, Corresponding PL intensity maps after the initial control cycle (\textbf{c}), and
after the subsequent live cycle with immersion in ethanol (\textbf{d}). Dashed line, outline of hBN flake. Inset of \textbf{c}:
Simultaneously acquired intensity map of the E$_{2g}$ Raman peak of hBN. \textbf{e}-\textbf{g}, Analogous characterizations of
an hBN sheet (thickness 12 nm) after control cycle and live cycle with immersion in anisole. \textbf{h}-\textbf{j},
Corresponding data set for an hBN sheet (thickness 12 nm) subjected to two subsequent control cycles, i.e., no exposure to
organic solvent. \textbf{k},\textbf{l}, Photograph (\textbf{k}) and PL intensity map (\textbf{l}) of an hBN sheet (thickness 20
nm) subjected only to a single live cycle with immersion in anisole. \textbf{m}, PL spectra from individual emitters, and
literature reference spectra (all at room temperature); eth-2, ethanol exposure in 2nd cycle (green circle in \textbf{d});
ani-2, anisole exposure in 2nd cycle (yellow circle in \textbf{g}); ani-1, anisole exposure in single cycle (cyan circle in
\textbf{l}); TRL, single terrylene molecule\cite{Kulzer1999a}; DBO, single dibenzo-ovalene molecule\cite{Chen2019}; SPEs in hBN
(red and orange lines\cite{Tran2016a,Tran2016b}). The similarity of spectral features points to PAH molecules as a common
origin of emission. $E_{0-0}$ and $E_{vibr}$ designate bands derived from the purely electronic and from vibronic transitions
of an individual molecule, respectively. All PL intensity maps were acquired at nominally identical conditions. Pairs of PL
intensity maps (first and second cycle) are shown with identical color scales.

\clearpage

\begin{figure}[ht]
\begin{center}
\includegraphics[width=9 cm]{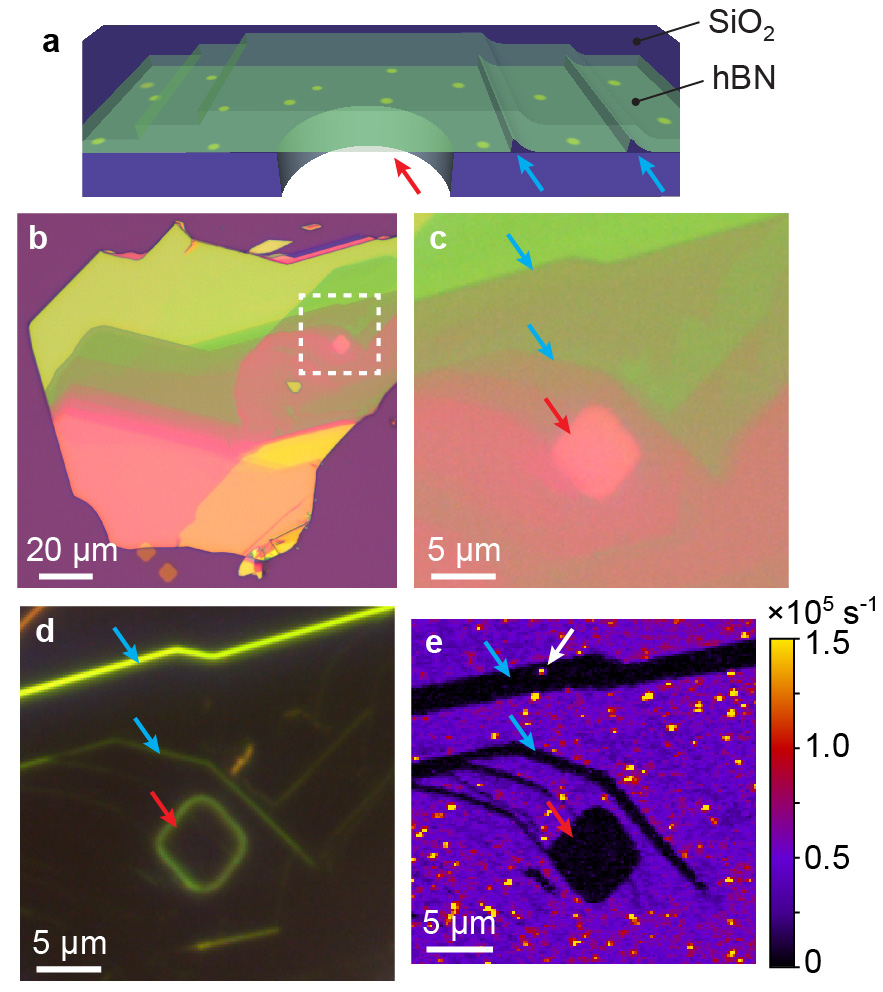}
\end{center}
\end{figure}

Figure 2: \textbf{Absence of emission centers in suspended hBN regions.} \textbf{a}, Illustration of a thick hBN flake with a
suspended area where the \ce{SiO2} substrate is recessed (red arrow). Bright points represent emission centers. \textbf{b},
Photograph of a thick hBN flake that contains a suspended area. \textbf{c}-\textbf{e}, Brightfield (\textbf{c}) and darkfield
(\textbf{d}) optical photographs, PL intensity map (\textbf{e}) of the area highlighted by a white box in \textbf{b}. In
addition to an area suspended over an alignment marker etched into the substrate (red arrows), the hBN flake contains
intrinsically suspended regions due to thickness steps at the flake's underside (blue arrows). Emission centers and homogeneous
PL background appear almost exclusively on regions supported by the substrate, and are absent in suspended hBN areas,
suggesting that the sources of PL are located at the hBN/\ce{SiO2} interface. Suspended area contains a single instance of
point-like emission (white arrow). The flake thickness is $\sim 260$ nm at the etched marker. Marker etch depth is 50 nm. The
sample underwent a single live cycle, involving exposure to anisole.

\clearpage

\begin{figure}[ht]
\begin{center}
\includegraphics[width=17 cm]{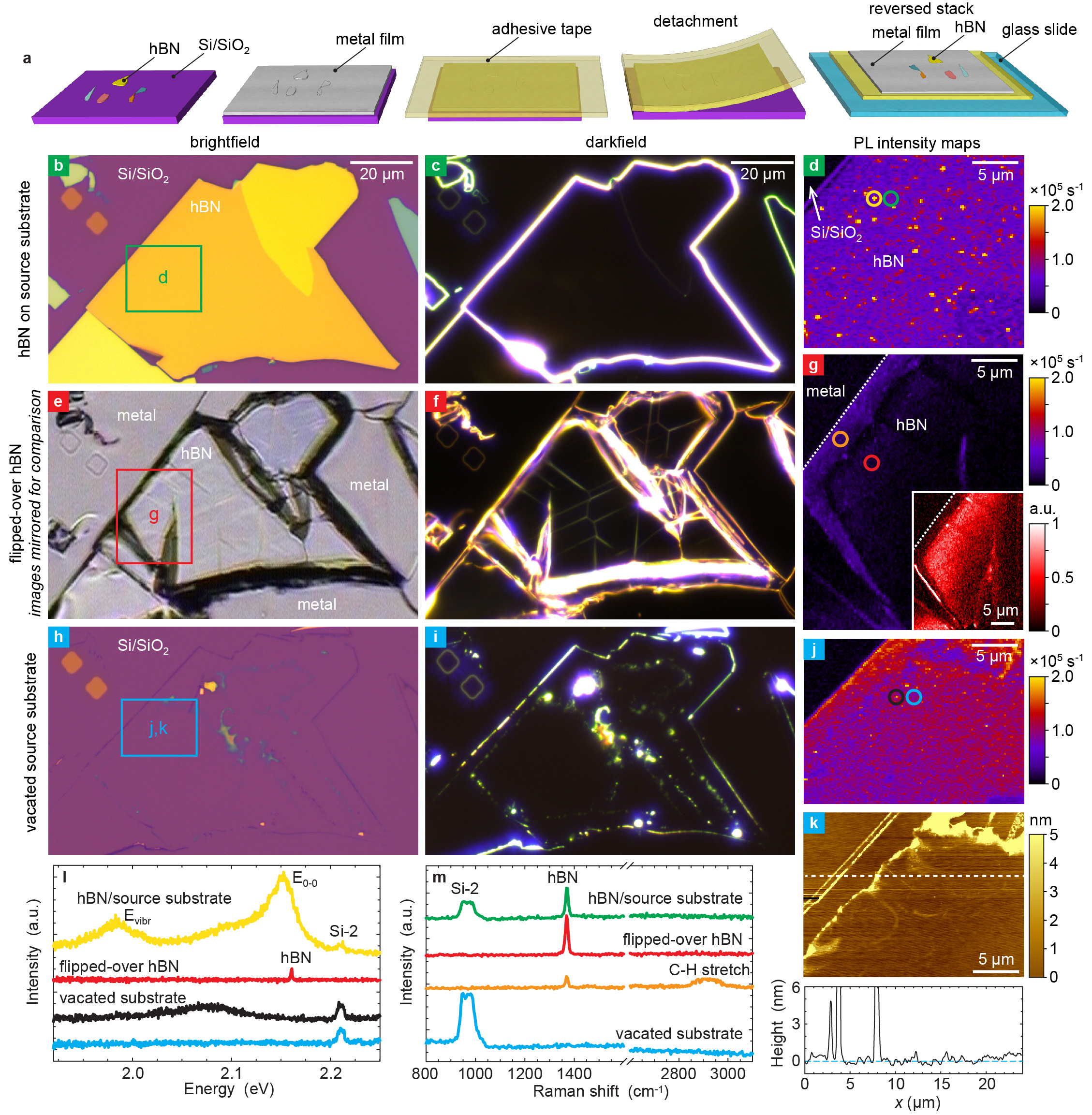}    
\end{center}
\end{figure}
\vspace{-0.5cm}
Figure 3: \textbf{Separating hBN flake from its substrate proves emitters to be located directly at the hBN/substrate
interface.} \textbf{a}, Schematic of hBN pickup and flipping-over procedure. Flakes and substrate are covered with a metal film
(5 nm Cr/70 nm Pt), and double-sided adhesive tape is attached to the metal film. The adhesive tape is lifted up, along with
attached metal film and hBN flakes. The tape is then reversed and attached to a microscope slide. \textbf{b},\textbf{c},
Brightfield (\textbf{b}) and darkfield (\textbf{c}) microscope images of a thick hBN flake (thickness 96 nm). \textbf{d}, PL
intensity map of the area indicated by green box in \textbf{b}. \textbf{e}-\textbf{g} Analogous characterizations of the same
hBN flake after pickup and reversal by means of the metal film. These three images have been mirrored horizontally for direct
comparison. Inset of \textbf{g}: Simultaneously acquired  intensity map of the E$_{2g}$ Raman peak of hBN.
\textbf{h}-\textbf{j}, Same characterizations of the vacated Si/\ce{SiO2} substrate. \textbf{k}, AFM topography of the vacated
substrate, same area as \textbf{j}. Bottom: Topography trace extracted at location indicated (dotted line), showing that no hBN
is left on the vacated substrate. \textbf{l},\textbf{m}, Representative PL emission (\textbf{l}) and Raman (\textbf{m}) spectra
from locations indicated in PL intensity maps (colored circles). The hBN flake discussed here underwent a single live cycle,
involving exposure to anisole. All microscope photos have identical lateral scale. All PL maps were collected with the same
measurement conditions and are shown with identical lateral and intensity scales. PL spectra correspond to individual pixels in
PL maps. Raman spectra were averaged over $7 \times 7$ pixels to improve signal/noise ratio. Spectra are offset vertically for
clarity. See Supplementary Fig. 11 for additional characterizations of this sample.

\clearpage

\begin{figure}[ht]
\begin{center}
\includegraphics[width=16 cm]{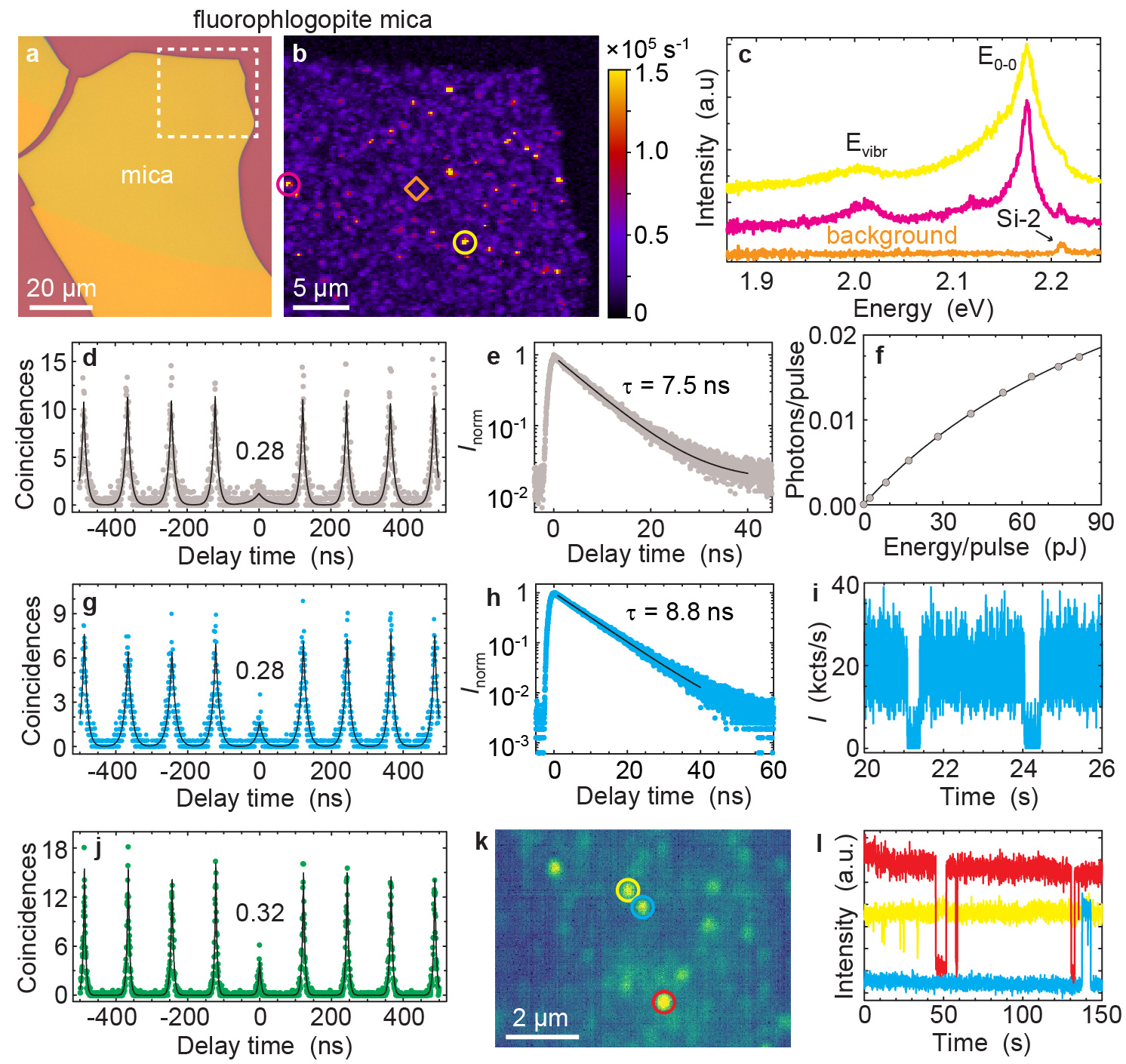}
\end{center}
\end{figure}
\vspace{-0.5cm}
Figure 4: \textbf{Single photon emission from fluorophlogopite mica.} \textbf{a}, Photograph of exfoliated mica flake
(thickness 115 nm). \textbf{b}, PL intensity map of area highlighted by a white box in \textbf{a}. \textbf{c}, PL spectra of
individual emission centers (circles indicated in \textbf{b}), and background spectrum (diamond symbol in \textbf{b}). Si-2,
second order Raman peak of the silicon substrate. \textbf{d-f}, Characterization of a SPE: photon correlation data
(\textbf{d}), time-resolved fluorescence intensity (\textbf{e}) with exponential fit function (black line), fluorescence
saturation curve (\textbf{f}) acquired with a pulsed laser (532 nm, 8.2 MHz), fit function $I = I_\infty \times P/(P+P_{sat})$
(black line) with $I_\infty = 0.046$ photons/pulse and $P_{sat} = 134$ pJ/pulse. The measured intensity $I = 0.017$
photons/pulse corresponds to $1.4 \times 10^5$ photons/s, limited by the excitation pulse separation of 122 ns. \textbf{g-i},
Characterization of another SPE: photon correlation data (\textbf{g}), time-resolved fluorescence intensity (\textbf{h}),
fluorescence intensity time trace (\textbf{i}). Binary blinking indicates single photon emission, consistent with $g^{(2)}(0) =
0.28$. \textbf{j}, Photon correlation data of a third SPE. \textbf{k}, Fluorescence intensity image of the mica sheet, acquired
under continuous wide-field laser illumination (532 nm). \textbf{l}, Intensity time traces of selected emitters (circles in
\textbf{k}). Binary on/off switching seen in two emitters is common for single photon emitters at room temperature. Following
exfoliation with adhesive tape, the sample underwent a protocol that included activation heat treatment at 850 \degC\ (see
Methods). PL spectra correspond to individual pixels in PL map. Spectra are offset vertically for clarity. Photon correlation,
time-resolved fluorescence and time trace data were acquired with a pulsed laser (532 nm, 12 $\mu$W, 8.2 MHz). All
characterizations were performed on the same mica sheet.

\end{document}